\documentclass[10pt,twocolumn,showpacs,longbibliography,amsmath,amssymb,osajnl,floatfix,superscriptaddress]{revtex4-1}
\usepackage{graphicx}
\usepackage{dcolumn}
\usepackage{bm}
\usepackage{xcolor}
\usepackage{txfonts}
\usepackage{microtype}
\usepackage{comment}
\usepackage[normalem]{ulem}

\begin{document}

\title{Generation of quasi-monoenergetic proton beams via quantum radiative compression}

\author{Feng Wan}
\affiliation{MOE Key Laboratory for Nonequilibrium Synthesis and Modulation of Condensed Matter, School of Physics, Xi'an Jiaotong University, Xi'an 710049, China}
\author{Wei-Quan Wang}
\affiliation{Department of Physics, National University of Defense Technology, Changsha, 410073, China}
\author{Qian Zhao}
\affiliation{MOE Key Laboratory for Nonequilibrium Synthesis and Modulation of Condensed Matter, School of Physics, Xi'an Jiaotong University, Xi'an 710049, China}
\author{Hao Zhang}
\affiliation{Department of Physics, National University of Defense Technology, Changsha, 410073, China}
\author{Tong-Pu Yu}\email{tongpu@nudt.edu.cn}
\affiliation{Department of Physics, National University of Defense Technology, Changsha, 410073, China}
\author{Wei-Min Wang}
\affiliation{Department of Physics and Beijing Key Laboratory of Opto-electronic Functional Materials and Micro–nano Devices, Renmin University of China, Beijing 100872, China}
\author{Wen-Chao Yan}
\affiliation{Key Laboratory for Laser Plasmas (MOE), School of Physics and Astronomy, Shanghai Jiao Tong University, Shanghai, 200240, China}\affiliation{Collaborative Innovation Center of IFSA (CICIFSA), Shanghai Jiao Tong University, Shanghai 200240, China}
\author{Yong-Tao Zhao}
\affiliation{MOE Key Laboratory for Nonequilibrium Synthesis and Modulation of Condensed Matter, School of Physics, Xi'an Jiaotong University, Xi'an 710049, China}
\author{Karen Z. Hatsagortsyan}
\affiliation{Max-Planck-Institut f\"{u}r Kernphysik, Saupfercheckweg 1,
69117 Heidelberg, Germany}
\author{Christoph H. Keitel}
\affiliation{Max-Planck-Institut f\"{u}r Kernphysik, Saupfercheckweg 1,
69117 Heidelberg, Germany}
\author{Sergei V. Bulanov}
\affiliation{Institute of Physics ASCR, v.v.i. (FZU), ELI BEAMLINES, Za Radnic\'i 835, Doln\'i B\v re\v zany, 252241, Czech Republic}
\affiliation{Kansai Photon Science Institute, National Institutes for Quantum and Radiological Science and Technology, 8-1-7 Umemidai, Kizugawa-shi, Kyoto, 619-0215, Japan}
\author{Jian-Xing Li}\email{jianxing@xjtu.edu.cn}
\affiliation{MOE Key Laboratory for Nonequilibrium Synthesis and Modulation of Condensed Matter, School of Physics, Xi'an Jiaotong University, Xi'an 710049, China}

\date{\today}

\begin{abstract}
Dense high-energy monoenergetic proton beams  are vital for wide applications,
	thus modern laser-plasma-based ion acceleration methods are aiming to obtain high-energy proton beams with energy spread as low as possible.
	In this work, we put forward a quantum radiative compression method to post-compress  a highly accelerated proton beam  and convert it to
a dense quasi-monoenergetic one.  We find that when the relativistic plasma produced by radiation pressure acceleration collides head-on with an ultraintense laser beam, large-amplitude plasma oscillations are excited due to quantum radiation-reaction and the ponderomotive force, which induce compression of the phase space of protons located in its acceleration phase with negative gradient.
Our three-dimensional spin-resolved QED particle-in-cell simulations show that  hollow-structure proton beams with a peak energy $\sim$ GeV, relative energy spread of  few percents and number $N_p\sim10^{10}$ (or $N_p\sim 10^9$ with a  $1\%$ energy spread) can be  produced  in near future
laser facilities, which may fulfill the requirements of important applications, such as, for radiography of ultra-thick dense materials, or as injectors of hadron colliders.
\end{abstract}

\maketitle
Laser-plasma-based ion acceleration can provide a much higher acceleration gradient (from GeV/m up to TeV/m), larger beam density ($\sim$ 1\% solid density) and shorter beam duration (fs-ps)
than conventional electrostatic or radio-frequency accelerators \cite{Mourou2006,Macchi2013a}. Benefited from the compactness and low expenses, these ion sources are
of paramount significance in broad applications, such as, material tomography \cite{Roth_2002,Borghesi_2004}, plasma radiography \cite{Borghesi_2002, Mackinnon_2006},  cancer therapy \cite{Bulanov_2002, Schardt2010, Bulanov_2014}, inertial confinement fusion (ICF) \cite{Roth_2001,Naumova_2009, Tikhonchuk_2010}, and nuclear physics \cite{Macchi2013}. Generally, they  require dense high-energy ion bunches with a rather low energy spread,
for instance, the energy spread of hundreds-of-MeV proton beam in the cancer therapy is limited to $\lesssim 1\%$ \cite{Macchi2013a}, that of hundreds-of-GeV
(up to TeV) proton beam in
the high precision experiments of high-energy physics is in the order of $10^{-4}$ \cite{LHC, RHIC}, and the resolution of  proton radiography highly relies on the proton flux, energy and spread \cite{Borghesi_2004}. Thus, high-flux high-energy monoenergetic proton beams are in great demand.

Recently, with rapid developments of ultraintense ultrashort laser techniques the peak intensities of modern laser pulses have achieved $I_0\sim 5\times 10^{22} {\rm W/cm^2}$ with pulse duration of tens of femtoseconds and energy fluctuation $\sim 1\%$ \cite{Yoon_2019_Achieving,Danson_2019_Petawatt,Gales_2018_extreme}.  Meanwhile, under-construction or upgrading laser facilities, (e.g. ELI-beamlines \cite{ELI}, SULF \cite{SULF}, Apollon \cite{Apollon}, etc.) are
aimed at intensities higher than $10^{23} {\rm W/cm^2}$.
With such intense lasers novel laser-plasma-based ion acceleration schemes attract  broad attention.
For instance, a hybrid scheme of radiation pressure sheath acceleration is experimentally demonstrated, achieving generation of proton beams with cutoff energy $\sim 100$ MeV, and  with an exponentially rolling off plateau in the spectrum \cite{Higginson2018}.  Collisionless shock acceleration can generate $\sim 10^5$ protons with  a peak energy $\sim 20$MeV, and energy spread $\sim 1\%$  \cite{Haberberger2011} (similarly, about $10^9$ protons of 9 MeV with an energy spread $\sim 30\%$  are obtained in
\cite{Zhang2017}), while hole-boring radiation pressure acceleration (RPA) at  $I_0\approx 10^{20} {\rm W/cm^2}$ can produce a maximal energy per nucleon  $\lesssim 30$MeV with energy spread $\sim 30\%$ \cite{Henig2009, Kar2012, Bin2015,Scullion2017}. By contrast, light-sail RPA~\cite{Esirkepov_2004}, due to the advantages in the energy conversion and scaling, could generate much higher-energy protons (typically $\sim$GeV; further acceleration to tens of GeV might be influenced and even interrupted because of the Rayleigh-Taylor-like instability)  with  energy spread of tens of percents in three-dimensional (3D) simulations (those in less realistic two-dimensional (2D) simulations could be narrower)
\cite{Pegoraro_2007, Macchi2005, Chen2009,Yu2010,Bulanov_2010,Ji_2014,Zhou2016,Wan2020a}.
 Radiation reaction in the classical regime \cite{Landau1975} is shown to improve the quality of RPA beams \cite{Tamburini2010,Tamburini2011}.
As is known, in light-sail RPA to obtain GeV and even higher-energy protons, the energy scaling law requires a high laser intensity of $I_0\gtrsim 10^{23} {\rm W/cm^2}$. In such an intense laser field though quantum radiation-reaction (QRR) effects \cite{Piazza2012}
will play a significant role in the plasma dynamics and must be taken into account \cite{Chen_2010,Capdessus2015,Wan_2019}. 
For instance, the proton energy spectra can be essentially disturbed by the stochastic nature of photon emission~\cite{Wan_2019}. Moreover, recent studies suggest that  the electron and photon polarization can reshape QRR and related plasma dynamics \cite{Seipt2018,Buescher2020,Xue2020}. Thus, the generation of  dense GeV monoenergetic   proton beams is still a great challenge.

\begin{figure}[t]
    \setlength{\abovecaptionskip}{-0cm}
	\includegraphics[width=\linewidth]{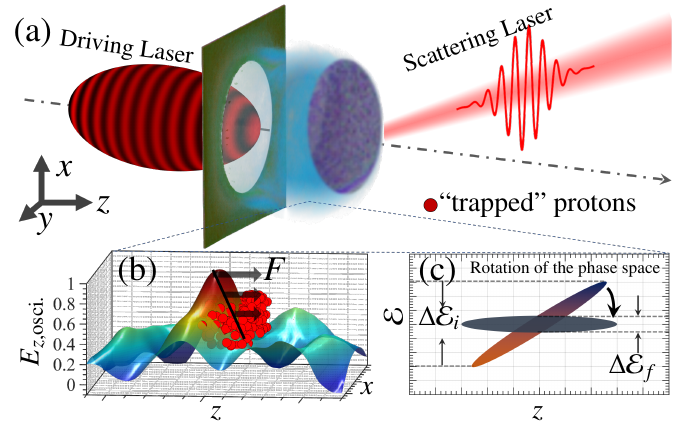}
	\caption{Interaction scenario. (a): The accelerated plasma via light-sail RPA by an intense CP driving laser collides with another intense LP scattering laser after the completed main acceleration phase.
(b): Protons are then ``trapped'' and further accelerated by the periodic oscillating longitudinal field $E_{z,{\rm osci.}}$, induced by  QRR effects and the ponderomotive force. The black-solid line and black arrows represent the negative gradient of $E_{z,{\rm osci.}}$ and the acceleration force $F$, respectively. Longer arrows denote larger $F$.  (c): The energy spread of protons is compressed  by $E_{z,{\rm osci.}}$ due to the rotation of their phase space.} \label{fig1}
\end{figure}

\begin{figure}[t]
	\includegraphics[width=\linewidth]{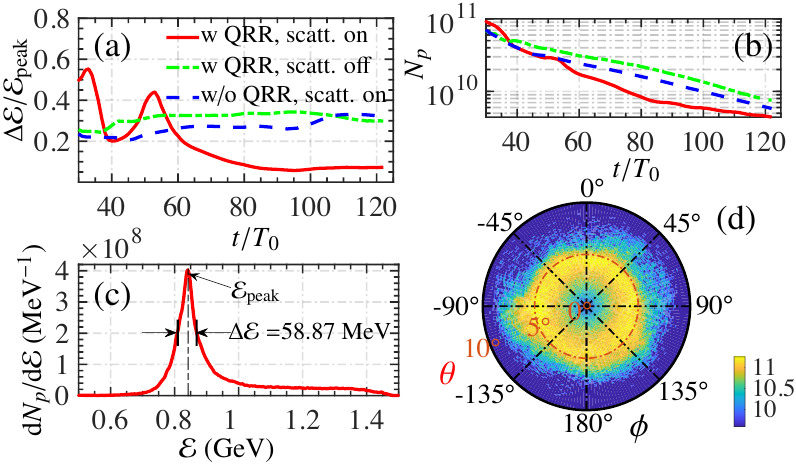}
	\caption{Generation of quasi-monoenergetic proton beam. (a) and (b): Time evolutions of the relative energy spread $\Delta\mathcal{E}/\mathcal{E}_{\rm peak}$ and number $N_p$ of the protons collected from the cylinder with a 7$\mu$m radius along the propagating axis of the driving laser. $\Delta\mathcal{E}$ is the absolute energy spread (FWHM) of the density spectrum [see (c)].   The red-solid, green-dash-dotted and blue-dashed curves indicate the cases including the QRR effects and  scattering laser, including QRR  but no scattering laser, and including scattering laser but no QRR, respectively. Here $t=0$ is redefined as the time of the driving laser reaching the left boundary of the simulation box (i.e. $t_1$ given in the text).  (c) and (d):
	For the case including QRR and  scattering laser, the energy spectrum  ${\rm d} N_p / {\rm d} \mathcal{E} ({\rm MeV}^{-1})$ and corresponding angular distribution $\log_{10}({\rm d^2}N_p/{\rm d} \theta {\rm d} \phi)$ of collected protons at $t=98 T_0$. $\theta$ and $\phi$ are the polar and azimuthal angles, respectively.
 Other laser and target parameters are given in the text.} \label{fig2}
\end{figure}

\begin{figure}[t]
	\includegraphics[width=\linewidth]{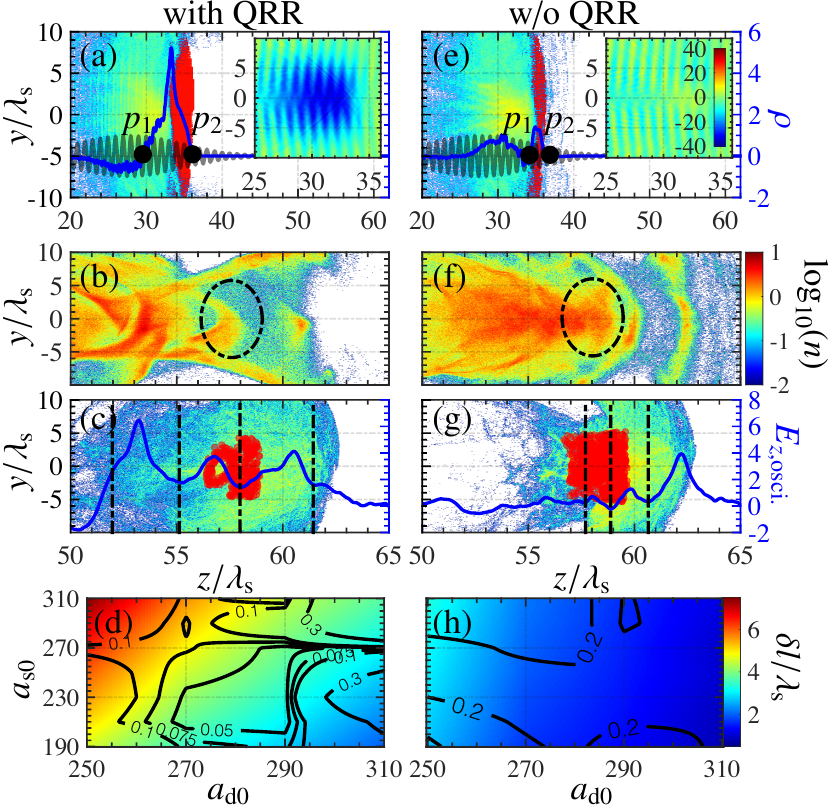}
	\caption{(a) and (e): Number density distributions of ions ($6n_C + n_p$, red) and electrons ($n_e$, colored background) at $t=66T_0$. Note that to show the relative slippage between electrons and ions, here ions are partially shaded by electrons. Solid-gray and solid-blue curves indicate
	 normalized scattering laser profile and charge density $\rho$ on the propagation axis ($y = 0$), respectively. Black dots $p_1$ and $p_2$ denote $\rho(p_1) = \rho(p_2) = 0$, and the relative slippage distance between the electron and ion layer is defined as $\delta l \equiv z(p_2) - z(p_1)$. Subfigures with color bar show the corresponding dimensionless longitudinal electric field $E_z$.  The corresponding electron energy spectrum is given in~\cite{supplemental}.
	 (b) and (f): Number density of electrons $n_e$ at $t=94T_0$. Black-dashed circles denote the positions of traced protons [shown in red in (c) and (g)]. (c) and (g): Number density of protons $n_p$ at $t = 94T_0$. Traced protons in FWHM of the proton energy spectra [i.e. $\Delta \mathcal{E}$; see Fig.~\ref{fig2}(c)] are labeled in red. Solid-blue curves denote $E_{z,{\rm osci.}}$ on the propagation axis ($y=0$), and the black-dashed curves are used to distinguish different field cycles.
(b)-(c) and (f)-(g) share the same color bar. In (a)-(c) and (e)-(g) $a_{\rm s0} = 300$.
(d) and (h): Correlation between  $\delta l$ (hot map) and  $\Delta\mathcal{E}/\mathcal{E}_{\rm peak}$ (contour) with respect to $a_{\rm d0}$ and $a_{\rm s0}$.
 Left and right columns show the cases including and excluding the QRR effects, respectively, and the  scattering laser is always included. Here 2D simulations are employed for simplicity.
  Other parameters are the same as those in Fig.~\ref{fig2}.} \label{fig3}
\end{figure}

\begin{figure}[t]
	\includegraphics[width=\linewidth]{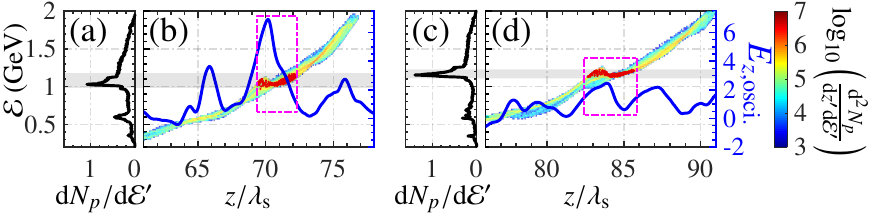}
	\caption{Simulations with QRR and scattering laser. (a) and (c): Energy spectra d$N_p$/d$\mathcal{E'}$ normalized by a factor of $9.39\times10^{16}$ vs $\mathcal{E}$ at $t = 114T_0$ and $t = 130T_0$, respectively. $\mathcal{E}' \equiv \mathcal{E}/{\rm GeV}$. (b) and (d): log$_{10}(\frac{{\rm d}^2N_p}{{\rm d}z'{\rm d}\mathcal{E'}})$ with respect to $z'=z/\lambda_{\rm s}$ and $\mathcal{E}$ at $t = 114T_0$ and $t = 130T_0$, respectively. The blue lines and red particles indicate $E_{z,{\rm osci.}} (y=0)$ and traced protons, respectively. The gray bands in (a)-(d) denote $\Delta\mathcal{E}$.
Other parameters are the same as those in Fig.~\ref{fig3}.} \label{fig4}
\end{figure}

In this Letter, we put forward a quantum radiative compression (QRC) method to generate dense GeV quasi-monoenergetic   proton beams [see the interaction scenario in Fig.~\ref{fig1}(a)].  In addition to the common light-sail RPA, when a circularly polarized (CP) laser pulse irradiates an ultra-thin target to generate and accelerate plasma, we apply another intense linearly polarized (LP) laser pulse head-on colliding with the accelerated plasma after RPA stage. In the second stage QRR dominates the plasma dynamics, inducing plasma oscillations with the assistance of the laser ponderomotive force.  Consequently, an oscillating longitudinal electric field $E_{z,{\rm osci.}}$ inside the plasma is excited  and further  accelerates and compresses protons to form a dense quasi-monoenergetic proton beam:
initially lower-energy protons experience a larger acceleration force of the plasma oscillations
$F$ 
to gain more energy and thus catch up initially higher-energy ones (resulting in a rotation of the phase space) [see  Figs.~\ref{fig1}(b) and (c) and detailed explanations in Figs.~\ref{fig3} and \ref{fig4}]. We underline that in the QRC process the scattering laser and QRR  are indispensable [see Fig.~\ref{fig2}].
To describe the plasma dynamics accurately in the applied QRR regime, 
we have implemented the spin-resolved Monte Carlo processes of electron dynamics and radiation 
\cite{li2019prl,Ligammaray_2020,Xue2020,Liu2020,Guo_2020} into the 3D particle-in-cell (PIC) code EPOCH \cite{Arber2015}.
With  up-coming laser facilities \cite{Yoon_2019_Achieving,Danson_2019_Petawatt, Gales_2018_extreme, ELI, SULF, Apollon},  
proton bunches with a peak energy of GeV order, relative energy spread of few percents and total number $N_p \sim 10^{10}$ can be obtained [see  Fig.~\ref{fig2}], to the benefit of many applications.

Sample results of generated dense GeV quasi-monoenergetic proton beams are illustrated in Fig.~\ref{fig2}. When choosing the laser and plasma parameters we ensure fulfilling the conditions for the relativistic transparency, light-sail RPA, and for the domination of QRR with respect to the laser ponderomotive force, as given in the following.
The peak intensity of the CP driving laser is $I_{\rm d0} \approx 8.56\times 10^{22} {\rm W/cm^2}$ (the corresponding invariant field parameter $a_{\rm d0} = eE_0/m_e c\omega_{\rm d} \simeq \sqrt{\frac{I_{\rm d0}}{1.37 \times 10^{18} {\rm W/cm^2}}} \lambda_{\rm d}[{\rm \mu m}] \approx 250$ with wavelength $\lambda_{\rm d} = 1{\rm \mu m}$). $-e$ and $m_e$ are the charge and mass of electron, respectively, $E_0$ and $\omega_{\rm d}$   the amplitude and  frequency of the driving laser, respectively, and $c$  the light speed in vacuum. The  profile of the driving laser pulse is $\frac{1}{2}\exp\left(-\frac{r^4}{w_{\rm d0}^4}\right) \{\tanh[2(t - t_1)] - \tanh[2(t - t_2)]\}$, with coordinate $r = \sqrt{x^2+y^2}$ and focal radius $w_{\rm d0} = 6 {\rm \mu m}$. $t_1 = 88T_0$ and $t_2 = 98T_0$ are the times of the front and tail of the driving laser entering the simulation box, respectively, with laser period $T_0$, and the pulse duration  $\tau_{\rm d} = t_2 - t_1$. We consider a fully ionized polystyrene target composed of $e^-$, $C^{6+}$ and $H^+$, with  number densities $n_e(e^{-}) = 300n_c$ and $n_p(H^+) = n_C(C^{6+}) = n_e(e^{-})/7$, and target thickness $l=0.3{\rm \mu m}$. $n_c = m_e\omega_{\rm d}^2/4\pi e^2$ is the critical plasma density. The laser and target parameters are optimized to meet the partially relativistic transparency condition $l/\lambda_{\rm d} \approx a_{\rm d0} n_c/\pi n_{e}\approx 0.3$ \cite{Macchi2009,Qiao2009}, which can suppress the target-deformation-induced instability \cite{Qiao2010}. For the LP scattering laser, the peak intensity is $I_{\rm s0} \approx 8.56 \times 10^{22} {\rm W/cm^2}$ ($a_{\rm s0} = 250$) with wavelength $\lambda_{\rm s} = \lambda_{\rm d}$, and the profile  $\frac{1}{2}\exp\left(-\frac{r^2}{w_{\rm s0}^2}\right) \{\tanh[2(t - t_3)] - \tanh[2(t - t_4)]\}$ with focal radius $w_{\rm s0} = 7{\rm \mu m}$. $t_3 = 3T_0$ and  $t_4 = 19T_0$ are the times of the front and tail of the scattering laser entering the simulation box, respectively, and the pulse duration   $\tau_{\rm s} = t_4 - t_3$. These synchronized two laser beams will be feasible soon by employing those multi-beam petawatt facilities or by splitting one laser beam in multi-petawatt laser
facilities (e.g., ELI-Beamlines, SULF and Apollon \cite{Yoon_2019_Achieving,Danson_2019_Petawatt, Gales_2018_extreme, ELI, SULF, Apollon}).
The 3D simulation box is placed at $-20{\rm \mu m} \leq z \leq 100{\rm \mu m}$, $-15{\rm \mu m} \leq x \leq 15{\rm \mu m}$ and   $-15{\rm \mu m} \leq y \leq 15{\rm \mu m}$, with mesh size $n_z \times n_x \times n_y = 6000 \times 300 \times 300$. The target is placed at $-18.65{\rm \mu m} \leq z \leq -18.35{\rm \mu m}$ and represented by 150 macro-electrons, 50 macro-protons and 10 macro-carbon ions per cell. In the following, the electromagnetic fields $E$ and $B$ are normalized by $ e/m_ec\omega_s$, the number density $n$ by $n_c$, and the charge density $\rho$ by $n_c e$.

For the given parameters, during light-sail RPA the electrons can be accelerated up to about 50 MeV with relativistic Lorentz factor $\gamma_e\approx 100$; see \cite{supplemental}.
Afterwards, they head-on collide with the scattering laser to emit photons via nonlinear Compton scattering.
The
 strong field quantum parameter in this setup is $\chi_e=(\hbar \omega_s/m_ec^2)\gamma_e a_{\rm s0}[1-v_e \rm cos(\theta_{coll.})/c]\sim 0.01-0.1$, which indicates that the QRR effects
are not negligible \cite{Piazza2012}, however, further electron-positron pair production via nonlinear Breit-Wheeler process is suppressed~\cite{Piazza2012,Wan_2020}. Here $v_e$ is the electron velocity and $\theta_{\rm coll.}$  the collision angle.
Our analysis in Figs.~\ref{fig3} and \ref{fig4} shows that indeed the electron motion is affected by QRR, and what is more remarkable, it has significant consequences for the ion dynamics due to the modification of charge separation forces.
Time evolutions of the relative energy spread $\Delta\mathcal{E}/\mathcal{E}_{\rm peak}$ and number $N_p$ of collected protons
along the propagation axis of the driving laser are shown in Figs.~\ref{fig2}(a) and (b). As including the scattering laser and QRR effects, before $t\approx 33T_0$ the relative energy spread $\Delta\mathcal{E}/\mathcal{E}_{\rm peak}$ first increases since the scattering laser pushes the electrons and consequently increases $\Delta\mathcal{E}$ by charge separation forces; in the range of $33T_0\lesssim t \lesssim 53 T_0$ instead $\Delta\mathcal{E}/\mathcal{E}_{\rm peak}$ is unstable because even though the scattering laser has left  the protons are not stably phase-matched with the excited oscillating longitudinal field $E_{z,{\rm osci.}}$ (see \cite{supplemental}); after $t\approx 53 T_0$  the $E_{z,{\rm osci.}}$ is becoming stable, and the protons are gradually  phase-matched with  $E_{z,{\rm osci.}}$
and  substantially compressed by $E_{z,{\rm osci.}}$ (negative gradient) from initial $\Delta\mathcal{E}/\mathcal{E}_{\rm peak}\gtrsim 40\%$ down to  $\Delta\mathcal{E}/\mathcal{E}_{\rm peak}\lesssim 6\%$, which keeps stable as $t\gtrsim 80T_0$ [see scenario in Figs.~\ref{fig1}(b) and (c) and physical reasons in Figs.~\ref{fig3} and \ref{fig4}].
The proton number $N_p$ is continuously reduced due to the transverse momenta.
On the contrary, as excluding the scattering laser
(common setup) or artificially removing the QRR effects (in which electron dynamics is
governed by the Lorentz force only,
 and the electron spin  by the Thomas-Bargmann-Michel-Telegdi equation~\cite{Thomas_1926, Thomas_1927, Bargmann_1959}), $\Delta\mathcal{E}/\mathcal{E}_{\rm peak}$ is always large ($\sim25\%-35\%$). Thus, QRR effects and the ponderomotive force, both provided by the scattering laser, result in the generation of a quasi-monoenergetic proton beam; see the specific energy spectrum and angular distribution in Figs.~\ref{fig2}(c) and (d).
 The final peak energy is $\mathcal{E}_{\rm peak} \approx 839$MeV with $\Delta \mathcal{E}/\mathcal{E}_{\rm peak} \approx 6 \%$,   total number  $N_p \approx 4.6\times 10^9$, and  radial angular divergence  $\theta\approx 68$mrad (about $3.9^\circ$). And, $N_p\approx 10^9$ within  $1\%$ energy spread at $\mathcal{E}_{\rm peak}$ can be obtained.
Such a proton beam may serve as an injector for a hadron collider
 or ion radiography source for ultra-thick targets \cite{King1999,Macchi2013a}. Moreover, the proton beam has a hollow structure  [see Fig.~\ref{fig2}(d)] which
may find an application as a high-energy positron collimator \cite{Wang2020}. Note that here the impact of spin effects on the proton dynamics is insignificant, in contrast to Refs.~\cite{Xue2020,Liu2020,Wan_2020}, since the employed target is not initially spin-polarized and the pair production  is negligible.

\begin{figure}[t]
	\includegraphics[width=\linewidth]{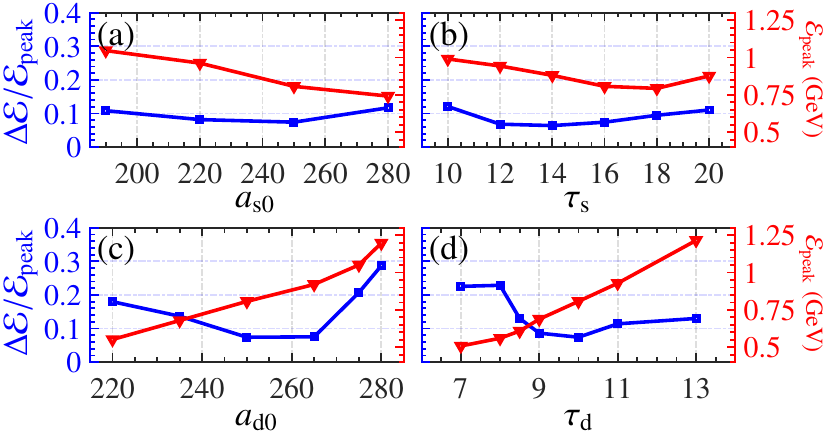}
	\caption{(a)-(d) Impact of $a_{\rm s0}$,  $\tau_{\rm s}$, $a_{\rm d0}$ and $\tau_{\rm d}$ on $\mathcal{E}_{\rm peak}$ and $\Delta\mathcal{E} / \mathcal{E}_{\rm peak}$, respectively. Other parameters are the same as those in Fig.~\ref{fig2}.} \label{fig5}
\end{figure}

The physics behind the proton QRC process
is analyzed in Figs.~\ref{fig3} and \ref{fig4}.
Due to the relativistic-transparency effects and Rayleigh-Taylor like instability, the driving laser will penetrate through the electron layer and induce an energy spread of accelerated protons~\cite{Pegoraro_2007}. Higher-energy protons move faster, and consequently, the target continues to expand \cite{Esirkepov_2004,Pegoraro_2007}. When the scattering laser head-on collides with the plasma, the electrons are decelerated by the scattering laser and  left behind the protons with a slippage distance $\delta l$, and the corresponding charge separation between electrons and protons along with the current flow can excite a strong longitudinal electric field $E_z = E_{z, {\rm stat.}} + E_{z, {\rm curr.}} < 0$ [see  Fig.~\ref{fig3}(a)]. Here the current-flow-induced $E_{z, {\rm curr.}} > 0$ is mainly derived from those protons moving faster than electrons, and $E_{z, {\rm curr.}}$ reduces $|E_z|$ (the charge separation field $E_{z, {\rm stat.}} < 0$).
When the scattering laser left, the electrons will be re-accelerated by $E_z < 0$ and excite a plasma oscillation $E_{z, {\rm osci.}}$ [see Figs.~\ref{fig3}(b) and (c)]. The protons synchronized with its negative-gradient phase  ($E_{z, {\rm osci.}} > 0, \partial_z E_{z, {\rm osci.}} < 0$) can be ``trapped''  and experience an energy-dependent acceleration field [see Fig.~\ref{fig3}(c) and the scenario in Figs.~\ref{fig1}(b) and (c)]. For instance, the protons locating near $z \approx 56.6 \lambda_{\rm s}$ are subjected to $E_{z, {\rm osci.}} \approx 4.0$, which is stronger than $E_{z, {\rm osci.}} \approx 1.7$ near $z \approx 58 \lambda_{\rm s}$. Since the proton energy  is approximately linearly proportional to its longitudinal position (i.e., lower-energy protons move relatively slower than higher-energy ones and thus are at the rear), $E_{z, {\rm osci.}}$ with negative-gradient can provide stronger acceleration for lower-energy protons near $z\approx 56.6\lambda_{\rm s}$. Consequently, the protons can be compressed in energy space to yield a low-energy-spread bunch [see Figs.~\ref{fig2} and \ref{fig4}].

Note that the compression efficiency is determined by the amplitude of  $E_{z, {\rm osci.}}$ and its spatial gradient $\partial_z E_{z, {\rm osci.}}$, which both rely on  $E_z$, and the oscillation wavelength is $\lambda_{\rm osci.} \sim \delta l$. Since $E_z = E_{z, {\rm stat.}} + E_{z, {\rm curr.}}$ and $E_{z, {\rm stat.}}$ is also proportional to $\delta l$, thus $\delta l$ plays a key role in the excitation of $E_{z, {\rm osci.}}$, and it is induced
by the ponderomotive force $F_{\rm p}$ and QRR force $F_{\rm QRR}$~\cite{supplemental, Avetissian2006}. Here, the relativistic ponderomotive force in $z$ direction is given by $F_{{\rm p}, z} \simeq -\frac{2}{\pi}(1+c/v_z) \partial_\eta a_s$ \cite{Lindman1977,Quesnel_1998,supplemental}, where $a_s(\eta)$ is  instantaneous invariant field parameter of the scattering laser with laser  phase $\eta=k_{\rm s}z-\omega_{\rm s} t$, and  wave vector $k_{\rm s}=2\pi/\lambda_{\rm s}$. For given parameters, $\max(F_{{\rm p}, z}) \approx \frac{4a_{\rm s0}}{\pi \tau_{\rm s}} \sqrt{\frac{2}{e'}} = 4.4$, where $e'$ is the natural logarithm base. While, the QRR force can be estimated via  $F_{\rm QRR} \simeq -\frac{2}{3}\chi_e^2 \alpha_f^2 \frac{\lambda_{\rm s}}{r_e}g(\chi_e)$~\cite{Niel2018,Baier1998,Bulanov2013a}, where $g(\chi_e) \simeq (1+8.93\chi_e + 2.41\chi_e^2)^{-2/3}$ is the quantum suppression function, $\alpha_f$ the fine structure constant and $r_e$  the classical electron radius. By averaging over a scattering laser period, $\langle F_{\rm QRR}\rangle \approx -\frac{4}{3\pi}\chi_e'^2 \alpha_f^2 \frac{\lambda_{\rm s}}{r_e}g(\chi_e')$, where $\chi_e' \approx \frac{2\hbar\omega_{\rm s}\gamma_e a_{\rm s0}}{m_ec^2}\exp\left(-\frac{\eta^2}{\tau_{\rm s}^2}\right)$. For $\langle \gamma_e \rangle \approx 70$, ${\rm max}\{\langle F_{\rm QRR}\rangle\} \approx 27$, which is much larger than $\max(F_{{\rm p}, z})$. Those two forces are balanced by
the excited longitudinal field: $\left\lvert E_z  \right\rvert= \left\lvert E_{z, {\rm curr.}} + E_{z, {\rm stat.}}\right\rvert \approx \left\lvert F_{\rm p,z} + F_{\rm QRR}\right\rvert$. From the simulation results [see $E_z$ in Fig.~\ref{fig3}(a)] ${\rm max}(|E_z|) \approx 40$ is very close to the estimated ${\rm max}(|F_{{\rm p},z} + F_{\rm QRR}|) \approx 32$.
We have also estimated ${\rm max}\left(E_{z, {\rm stat.}}\right) = E_{z, {\rm stat.}}(p_1) = -\int_{z(p_2)}^{z(p_1)}\rho(z){\rm d}z \simeq \frac{\rho_{\rm max} \delta l}{2} \approx -20\pi$ with $\rho_{\rm max} \approx 4$ and $\delta l \approx 5\lambda_{\rm s}$, which is in fact larger than $\left\lvert E_{z} \right\rvert$ due to counteracting $E_{z, {\rm curr.}} > 0$.
The QRC is caused by the further excited oscillation field  $E_{z, {\rm osci.}} \approx 4 - 5$ with $\lambda_{\rm osci.} \approx 3\lambda_{\rm s}$ [see Fig.~\ref{fig3}(c)]. The phase space of those ``trapped'' protons continuously rotates [see Figs.~\ref{fig4}(b) and (d)], and the energy spread is reduced approximately from 200 MeV at $t = 114T_0$ to 100MeV at $t = 130T_0$ [see Figs.~\ref{fig4}(a) and (c)]. This compression effects can sustain about 10-30 periods, and the final energy spread can be compressed down to few percents. The correlation between the final energy spread $\Delta \mathcal{E}/\mathcal{E}_{\rm peak}$ and the slippage $\delta l$ is illustrated in Fig.~\ref{fig3}(d).

As given in the applied condition $F_{\rm QRR}\gg F_{{\rm p},z}$, the key role of QRR for the QRC process is clear, which is described by the QRC parameter $|F_{\rm QRR}/ F_{{\rm p},z}|\sim {\cal R}\equiv (\chi^2\lambda_s\tau_s/a_{0s})(\alpha^2 /r_e)$. In fact,
as we artificially remove QRR, the compression effects will be greatly suppressed [see Figs.~\ref{fig3}(e)-(h)].  In this case $E_{z, {\rm peak}} \approx {\rm max} \left(F_{{\rm p},z}\right) \approx 5$ and $\delta l \approx 3\lambda_{\rm s}$ [see Fig.~\ref{fig3}(e)].
 $E_{z, {\rm osci.}} \approx 1.0$ with $\lambda_{\rm osci.} \approx \lambda_{\rm s}$ [see Fig.~\ref{fig3}(g)]. And the final compression effects are rather weak [see the correlation between $\Delta \mathcal{E} / \mathcal{E}_{\rm peak}$ and $\delta l$ in Fig.~\ref{fig3}(h)].

For the experimental feasibility, we investigate the impact of the driving and scattering laser parameters  on the QRC efficiency, as shown in Fig.~\ref{fig5}.
As $190 \le a_{\rm s0} \le 280$, max($\chi_e$) $\approx 0.05 \sim 0.07$ and $F_{\rm QRR} \approx 26.8 - 33.1$, which is still much larger than $F_{\rm p,z}$, and  QRC takes place. While at smaller $a_{\rm s0}$ the QRC is suppressed, at larger $a_{\rm s0}$ the QRC will be enhanced up to the point when $\delta l$ is comparable to the longitudinal thickness of the plasma.  In deep quantum regime with $\chi_e(\propto a_{\rm s0}) \gg1$, the quantum stochasticity also will increase the energy spread of the electrons and further that of the protons \cite{Neitz_2013}. From the parameter ${\cal R}$ we can deduce that $\tau_{\rm s}$ plays a similar role as $a_{\rm s0}$ [see Figs.~\ref{fig5}(a) and (b)].
Moreover, the driving laser determines the energies of electrons and protons in light-sail RPA and therefore affects many parameters, e.g., $\chi_e$, $n_e$ and $n_i$.
Our 3D simulations show that to obtain $\Delta\mathcal{E}/\mathcal{E}_{\rm peak} \lesssim 10\%$, the optimal driving laser intensity is  $240\le a_{\rm d0} \le 270$ and the pulse duration is  $8.5 \le \tau_{\rm d} \lesssim 13$ [see Figs.~\ref{fig5}(c) and (d)]. As expected, $\mathcal{E}_{\rm peak}$ is proportional to $a_{\rm d0}$ and $\tau_{\rm d}$. However, when $a_{\rm d0}$ is much lower than the transparency condition, i.e., $a_{\rm d0} \ll \frac{\pi n_e l}{n_c \lambda_{\rm d}}$~\cite{Vshivkov_1998, Bulanov_2016}, $\gamma_e$ and $\chi_e$ will be much smaller, and consequently $F_{\rm QRR}$ is rather weak and the QRC will be ineffective. On the contrary, if $a_{\rm d0}$ is too high, the target deformation will be much earlier, which will also limit the effective acceleration.  $\tau_{\rm d}$ has  similar effects:  a too short driving laser can not effectively accelerate plasma [e.g. $\tau_{\rm d} \lesssim 9T_0$ in Fig.~\ref{fig5}(d)], while a too long driving laser will generate very high-energy protons, which are hard to be compressed due to limited $E_{z, {\rm osci.}}$ and $\lambda_{\rm osci.}$. Note that these results are collected at $t=130T_0$, and for longer pulses the energy spectra may be further compressed when extending the simulation sizes.

In conclusion, we have proposed the QRC method  to generate dense GeV quasi-monoenergetic proton beams, which is based on  QRR effects  for the proton dynamics in plasma. With up-coming laser facilities, such as ELI-beamlines, SULF, Appolon, our 3D spin-resolved QED-PIC simulations show that  hollow-structure proton beams with peak energy $\mathcal{E}_{\rm peak}\sim$ GeV, energy spread $\lesssim$ 6\% and number $N_p \sim 10^{10}$ ($N_p \sim 10^{9}$ within $\Delta \mathcal{E} / \mathcal{E}_{\rm peak} \leq 1\%$) can be obtained, which may fulfill the requirements of high-resolution proton imaging, high-energy particle physics and relativistic positron collimation.\\

{\it Acknowledgment:} This work is supported by the National Key Research and Development Program of China (Grant Nos. 2018YFA0404801, 2018YFA0404802), the National Natural Science Foundation of China (Grant Nos. 11875319, 12022506, 12005298, 11874295, 11804269, U1532263), and the Research Project of NUDT (ZK18-02-02; ZK19-25). The work is also supported by the project High Field Initiative
(CZ.02.1.01/0.0/0.0/15\_003/0000449)
from the European Regional Development Fund.

\bibliography{lib}

\begin{thebibliography}{71}%
\makeatletter
\providecommand \@ifxundefined [1]{%
 \@ifx{#1\undefined}
}%
\providecommand \@ifnum [1]{%
 \ifnum #1\expandafter \@firstoftwo
 \else \expandafter \@secondoftwo
 \fi
}%
\providecommand \@ifx [1]{%
 \ifx #1\expandafter \@firstoftwo
 \else \expandafter \@secondoftwo
 \fi
}%
\providecommand \natexlab [1]{#1}%
\providecommand \enquote  [1]{``#1''}%
\providecommand \bibnamefont  [1]{#1}%
\providecommand \bibfnamefont [1]{#1}%
\providecommand \citenamefont [1]{#1}%
\providecommand \href@noop [0]{\@secondoftwo}%
\providecommand \href [0]{\begingroup \@sanitize@url \@href}%
\providecommand \@href[1]{\@@startlink{#1}\@@href}%
\providecommand \@@href[1]{\endgroup#1\@@endlink}%
\providecommand \@sanitize@url [0]{\catcode `\\12\catcode `\$12\catcode
  `\&12\catcode `\#12\catcode `\^12\catcode `\_12\catcode `\%12\relax}%
\providecommand \@@startlink[1]{}%
\providecommand \@@endlink[0]{}%
\providecommand \url  [0]{\begingroup\@sanitize@url \@url }%
\providecommand \@url [1]{\endgroup\@href {#1}{\urlprefix }}%
\providecommand \urlprefix  [0]{URL }%
\providecommand \Eprint [0]{\href }%
\providecommand \doibase [0]{http://dx.doi.org/}%
\providecommand \selectlanguage [0]{\@gobble}%
\providecommand \bibinfo  [0]{\@secondoftwo}%
\providecommand \bibfield  [0]{\@secondoftwo}%
\providecommand \translation [1]{[#1]}%
\providecommand \BibitemOpen [0]{}%
\providecommand \bibitemStop [0]{}%
\providecommand \bibitemNoStop [0]{.\EOS\space}%
\providecommand \EOS [0]{\spacefactor3000\relax}%
\providecommand \BibitemShut  [1]{\csname bibitem#1\endcsname}%
\let\auto@bib@innerbib\@empty
\bibitem [{\citenamefont {Mourou}\ \emph {et~al.}(2006)\citenamefont {Mourou},
  \citenamefont {Tajima},\ and\ \citenamefont {Bulanov}}]{Mourou2006}%
  \BibitemOpen
  \bibfield  {author} {\bibinfo {author} {\bibfnamefont {Gerard~A.}\
  \bibnamefont {Mourou}}, \bibinfo {author} {\bibfnamefont {Toshiki}\
  \bibnamefont {Tajima}}, \ and\ \bibinfo {author} {\bibfnamefont {Sergei~V.}\
  \bibnamefont {Bulanov}},\ }\bibfield  {title} {\enquote {\bibinfo {title}
  {Optics in the relativistic regime},}\ }\href {\doibase
  10.1103/revmodphys.78.309} {\bibfield  {journal} {\bibinfo  {journal} {Rev.
  Mod. Phys.}\ }\textbf {\bibinfo {volume} {78}},\ \bibinfo {pages} {309--371}
  (\bibinfo {year} {2006})}\BibitemShut {NoStop}%
\bibitem [{\citenamefont {Macchi}\ \emph
  {et~al.}(2013{\natexlab{a}})\citenamefont {Macchi}, \citenamefont
  {Borghesi},\ and\ \citenamefont {Passoni}}]{Macchi2013a}%
  \BibitemOpen
  \bibfield  {author} {\bibinfo {author} {\bibfnamefont {Andrea}\ \bibnamefont
  {Macchi}}, \bibinfo {author} {\bibfnamefont {Marco}\ \bibnamefont
  {Borghesi}}, \ and\ \bibinfo {author} {\bibfnamefont {Matteo}\ \bibnamefont
  {Passoni}},\ }\bibfield  {title} {\enquote {\bibinfo {title} {Ion
  acceleration by superintense laser-plasma interaction},}\ }\href {\doibase
  10.1103/revmodphys.85.751} {\bibfield  {journal} {\bibinfo  {journal} {Rev.
  Mod. Phys.}\ }\textbf {\bibinfo {volume} {85}},\ \bibinfo {pages} {751--793}
  (\bibinfo {year} {2013}{\natexlab{a}})}\BibitemShut {NoStop}%
\bibitem [{\citenamefont {Roth}\ \emph {et~al.}(2002)\citenamefont {Roth},
  \citenamefont {Blazevic}, \citenamefont {Geissel}, \citenamefont {Schlegel},
  \citenamefont {Cowan}, \citenamefont {Allen}, \citenamefont {Gauthier},
  \citenamefont {Audebert}, \citenamefont {Fuchs}, \citenamefont {ter Vehn},
  \citenamefont {Hegelich}, \citenamefont {Karsch},\ and\ \citenamefont
  {Pukhov}}]{Roth_2002}%
  \BibitemOpen
  \bibfield  {author} {\bibinfo {author} {\bibfnamefont {M.}~\bibnamefont
  {Roth}}, \bibinfo {author} {\bibfnamefont {A.}~\bibnamefont {Blazevic}},
  \bibinfo {author} {\bibfnamefont {M.}~\bibnamefont {Geissel}}, \bibinfo
  {author} {\bibfnamefont {T.}~\bibnamefont {Schlegel}}, \bibinfo {author}
  {\bibfnamefont {T.~E.}\ \bibnamefont {Cowan}}, \bibinfo {author}
  {\bibfnamefont {M.}~\bibnamefont {Allen}}, \bibinfo {author} {\bibfnamefont
  {J.-C.}\ \bibnamefont {Gauthier}}, \bibinfo {author} {\bibfnamefont
  {P.}~\bibnamefont {Audebert}}, \bibinfo {author} {\bibfnamefont
  {J.}~\bibnamefont {Fuchs}}, \bibinfo {author} {\bibfnamefont {J.~Meyer}\
  \bibnamefont {ter Vehn}}, \bibinfo {author} {\bibfnamefont {M.}~\bibnamefont
  {Hegelich}}, \bibinfo {author} {\bibfnamefont {S.}~\bibnamefont {Karsch}}, \
  and\ \bibinfo {author} {\bibfnamefont {A.}~\bibnamefont {Pukhov}},\
  }\bibfield  {title} {\enquote {\bibinfo {title} {Energetic ions generated by
  laser pulses: A detailed study on target properties},}\ }\href {\doibase
  10.1103/PhysRevSTAB.5.061301} {\bibfield  {journal} {\bibinfo  {journal}
  {Phys. Rev. ST Accel. Beams}\ }\textbf {\bibinfo {volume} {5}},\ \bibinfo
  {pages} {061301} (\bibinfo {year} {2002})}\BibitemShut {NoStop}%
\bibitem [{\citenamefont {Borghesi}\ \emph {et~al.}(2004)\citenamefont
  {Borghesi}, \citenamefont {Mackinnon}, \citenamefont {Campbell},
  \citenamefont {Hicks}, \citenamefont {Kar}, \citenamefont {Patel},
  \citenamefont {Price}, \citenamefont {Romagnani}, \citenamefont {Schiavi},\
  and\ \citenamefont {Willi}}]{Borghesi_2004}%
  \BibitemOpen
  \bibfield  {author} {\bibinfo {author} {\bibfnamefont {M.}~\bibnamefont
  {Borghesi}}, \bibinfo {author} {\bibfnamefont {A.~J.}\ \bibnamefont
  {Mackinnon}}, \bibinfo {author} {\bibfnamefont {D.~H.}\ \bibnamefont
  {Campbell}}, \bibinfo {author} {\bibfnamefont {D.~G.}\ \bibnamefont {Hicks}},
  \bibinfo {author} {\bibfnamefont {S.}~\bibnamefont {Kar}}, \bibinfo {author}
  {\bibfnamefont {P.~K.}\ \bibnamefont {Patel}}, \bibinfo {author}
  {\bibfnamefont {D.}~\bibnamefont {Price}}, \bibinfo {author} {\bibfnamefont
  {L.}~\bibnamefont {Romagnani}}, \bibinfo {author} {\bibfnamefont
  {A.}~\bibnamefont {Schiavi}}, \ and\ \bibinfo {author} {\bibfnamefont
  {O.}~\bibnamefont {Willi}},\ }\bibfield  {title} {\enquote {\bibinfo {title}
  {Multi-{MeV} proton source investigations in ultraintense laser-foil
  interactions},}\ }\href {\doibase 10.1103/PhysRevLett.92.055003} {\bibfield
  {journal} {\bibinfo  {journal} {Phys. Rev. Lett.}\ }\textbf {\bibinfo
  {volume} {92}},\ \bibinfo {pages} {055003} (\bibinfo {year}
  {2004})}\BibitemShut {NoStop}%
\bibitem [{\citenamefont {Borghesi}\ \emph {et~al.}(2002)\citenamefont
  {Borghesi}, \citenamefont {Campbell}, \citenamefont {Schiavi}, \citenamefont
  {Haines}, \citenamefont {Willi}, \citenamefont {MacKinnon}, \citenamefont
  {Patel}, \citenamefont {Gizzi}, \citenamefont {Galimberti}, \citenamefont
  {Clarke}, \citenamefont {Pegoraro}, \citenamefont {Ruhl},\ and\ \citenamefont
  {Bulanov}}]{Borghesi_2002}%
  \BibitemOpen
  \bibfield  {author} {\bibinfo {author} {\bibfnamefont {M.}~\bibnamefont
  {Borghesi}}, \bibinfo {author} {\bibfnamefont {D.~H.}\ \bibnamefont
  {Campbell}}, \bibinfo {author} {\bibfnamefont {A.}~\bibnamefont {Schiavi}},
  \bibinfo {author} {\bibfnamefont {M.~G.}\ \bibnamefont {Haines}}, \bibinfo
  {author} {\bibfnamefont {O.}~\bibnamefont {Willi}}, \bibinfo {author}
  {\bibfnamefont {A.~J.}\ \bibnamefont {MacKinnon}}, \bibinfo {author}
  {\bibfnamefont {P.}~\bibnamefont {Patel}}, \bibinfo {author} {\bibfnamefont
  {L.~A.}\ \bibnamefont {Gizzi}}, \bibinfo {author} {\bibfnamefont
  {M.}~\bibnamefont {Galimberti}}, \bibinfo {author} {\bibfnamefont {R.~J.}\
  \bibnamefont {Clarke}}, \bibinfo {author} {\bibfnamefont {F.}~\bibnamefont
  {Pegoraro}}, \bibinfo {author} {\bibfnamefont {H.}~\bibnamefont {Ruhl}}, \
  and\ \bibinfo {author} {\bibfnamefont {S.}~\bibnamefont {Bulanov}},\
  }\bibfield  {title} {\enquote {\bibinfo {title} {Electric field detection in
  laser-plasma interaction experiments via the proton imaging technique},}\
  }\href {\doibase 10.1063/1.1459457} {\bibfield  {journal} {\bibinfo
  {journal} {Phys. Plasmas}\ }\textbf {\bibinfo {volume} {9}},\ \bibinfo
  {pages} {2214--2220} (\bibinfo {year} {2002})}\BibitemShut {NoStop}%
\bibitem [{\citenamefont {Mackinnon}\ \emph {et~al.}(2006)\citenamefont
  {Mackinnon}, \citenamefont {Patel}, \citenamefont {Borghesi}, \citenamefont
  {Clarke}, \citenamefont {Freeman}, \citenamefont {Habara}, \citenamefont
  {Hatchett}, \citenamefont {Hey}, \citenamefont {Hicks}, \citenamefont {Kar},
  \citenamefont {Key}, \citenamefont {King}, \citenamefont {Lancaster},
  \citenamefont {Neely}, \citenamefont {Nikkro}, \citenamefont {Norreys},
  \citenamefont {Notley}, \citenamefont {Phillips}, \citenamefont {Romagnani},
  \citenamefont {Snavely}, \citenamefont {Stephens},\ and\ \citenamefont
  {Town}}]{Mackinnon_2006}%
  \BibitemOpen
  \bibfield  {author} {\bibinfo {author} {\bibfnamefont {A.~J.}\ \bibnamefont
  {Mackinnon}}, \bibinfo {author} {\bibfnamefont {P.~K.}\ \bibnamefont
  {Patel}}, \bibinfo {author} {\bibfnamefont {M.}~\bibnamefont {Borghesi}},
  \bibinfo {author} {\bibfnamefont {R.~C.}\ \bibnamefont {Clarke}}, \bibinfo
  {author} {\bibfnamefont {R.~R.}\ \bibnamefont {Freeman}}, \bibinfo {author}
  {\bibfnamefont {H.}~\bibnamefont {Habara}}, \bibinfo {author} {\bibfnamefont
  {S.~P.}\ \bibnamefont {Hatchett}}, \bibinfo {author} {\bibfnamefont
  {D.}~\bibnamefont {Hey}}, \bibinfo {author} {\bibfnamefont {D.~G.}\
  \bibnamefont {Hicks}}, \bibinfo {author} {\bibfnamefont {S.}~\bibnamefont
  {Kar}}, \bibinfo {author} {\bibfnamefont {M.~H.}\ \bibnamefont {Key}},
  \bibinfo {author} {\bibfnamefont {J.~A.}\ \bibnamefont {King}}, \bibinfo
  {author} {\bibfnamefont {K.}~\bibnamefont {Lancaster}}, \bibinfo {author}
  {\bibfnamefont {D.}~\bibnamefont {Neely}}, \bibinfo {author} {\bibfnamefont
  {A.}~\bibnamefont {Nikkro}}, \bibinfo {author} {\bibfnamefont {P.~A.}\
  \bibnamefont {Norreys}}, \bibinfo {author} {\bibfnamefont {M.~M.}\
  \bibnamefont {Notley}}, \bibinfo {author} {\bibfnamefont {T.~W.}\
  \bibnamefont {Phillips}}, \bibinfo {author} {\bibfnamefont {L.}~\bibnamefont
  {Romagnani}}, \bibinfo {author} {\bibfnamefont {R.~A.}\ \bibnamefont
  {Snavely}}, \bibinfo {author} {\bibfnamefont {R.~B.}\ \bibnamefont
  {Stephens}}, \ and\ \bibinfo {author} {\bibfnamefont {R.~P.~J.}\ \bibnamefont
  {Town}},\ }\bibfield  {title} {\enquote {\bibinfo {title} {Proton radiography
  of a laser-driven implosion},}\ }\href {\doibase
  10.1103/PhysRevLett.97.045001} {\bibfield  {journal} {\bibinfo  {journal}
  {Phys. Rev. Lett.}\ }\textbf {\bibinfo {volume} {97}},\ \bibinfo {pages}
  {045001} (\bibinfo {year} {2006})}\BibitemShut {NoStop}%
\bibitem [{\citenamefont {Bulanov}\ and\ \citenamefont
  {Khoroshkov}(2002)}]{Bulanov_2002}%
  \BibitemOpen
  \bibfield  {author} {\bibinfo {author} {\bibfnamefont {S.~V.}\ \bibnamefont
  {Bulanov}}\ and\ \bibinfo {author} {\bibfnamefont {V.~S.}\ \bibnamefont
  {Khoroshkov}},\ }\bibfield  {title} {\enquote {\bibinfo {title} {Feasibility
  of using laser ion accelerators in proton therapy},}\ }\href@noop {}
  {\bibfield  {journal} {\bibinfo  {journal} {Plasma Phys. Rep.}\ }\textbf
  {\bibinfo {volume} {28}},\ \bibinfo {pages} {453} (\bibinfo {year}
  {2002})}\BibitemShut {NoStop}%
\bibitem [{\citenamefont {Schardt}\ \emph {et~al.}(2010)\citenamefont
  {Schardt}, \citenamefont {Elsässer},\ and\ \citenamefont
  {Schulz-Ertner}}]{Schardt2010}%
  \BibitemOpen
  \bibfield  {author} {\bibinfo {author} {\bibfnamefont {Dieter}\ \bibnamefont
  {Schardt}}, \bibinfo {author} {\bibfnamefont {Thilo}\ \bibnamefont
  {Elsässer}}, \ and\ \bibinfo {author} {\bibfnamefont {Daniela}\ \bibnamefont
  {Schulz-Ertner}},\ }\bibfield  {title} {\enquote {\bibinfo {title} {Heavy-ion
  tumor therapy: Physical and radiobiological benefits},}\ }\href {\doibase
  10.1103/RevModPhys.82.383} {\bibfield  {journal} {\bibinfo  {journal} {Rev.
  Mod. Phys.}\ }\textbf {\bibinfo {volume} {82}},\ \bibinfo {pages} {383--425}
  (\bibinfo {year} {2010})}\BibitemShut {NoStop}%
\bibitem [{\citenamefont {Bulanov}\ \emph {et~al.}(2014)\citenamefont
  {Bulanov}, \citenamefont {Wilkens}, \citenamefont {Molls}, \citenamefont
  {Esirkepov}, \citenamefont {Korn}, \citenamefont {Kraft}, \citenamefont
  {Kraft},\ and\ \citenamefont {Khoroshkov}}]{Bulanov_2014}%
  \BibitemOpen
  \bibfield  {author} {\bibinfo {author} {\bibfnamefont {S.~V.}\ \bibnamefont
  {Bulanov}}, \bibinfo {author} {\bibfnamefont {J.~J.}\ \bibnamefont
  {Wilkens}}, \bibinfo {author} {\bibfnamefont {M.}~\bibnamefont {Molls}},
  \bibinfo {author} {\bibfnamefont {T.~Zh.}\ \bibnamefont {Esirkepov}},
  \bibinfo {author} {\bibfnamefont {G.}~\bibnamefont {Korn}}, \bibinfo {author}
  {\bibfnamefont {G.}~\bibnamefont {Kraft}}, \bibinfo {author} {\bibfnamefont
  {S.~D.}\ \bibnamefont {Kraft}}, \ and\ \bibinfo {author} {\bibfnamefont
  {V.~S.}\ \bibnamefont {Khoroshkov}},\ }\bibfield  {title} {\enquote {\bibinfo
  {title} {Laser ion acceleration for hadron therapy},}\ }\href@noop {}
  {\bibfield  {journal} {\bibinfo  {journal} {Physics Uspekhi}\ }\textbf
  {\bibinfo {volume} {57}},\ \bibinfo {pages} {1149} (\bibinfo {year}
  {2014})}\BibitemShut {NoStop}%
\bibitem [{\citenamefont {Roth}\ \emph {et~al.}(2001)\citenamefont {Roth},
  \citenamefont {Cowan}, \citenamefont {Key}, \citenamefont {Hatchett},
  \citenamefont {Brown}, \citenamefont {Fountain}, \citenamefont {Johnson},
  \citenamefont {Pennington}, \citenamefont {Snavely}, \citenamefont {Wilks},
  \citenamefont {Yasuike}, \citenamefont {Ruhl}, \citenamefont {Pegoraro},
  \citenamefont {Bulanov}, \citenamefont {Campbell}, \citenamefont {Perry},\
  and\ \citenamefont {Powell}}]{Roth_2001}%
  \BibitemOpen
  \bibfield  {author} {\bibinfo {author} {\bibfnamefont {M.}~\bibnamefont
  {Roth}}, \bibinfo {author} {\bibfnamefont {T.~E.}\ \bibnamefont {Cowan}},
  \bibinfo {author} {\bibfnamefont {M.~H.}\ \bibnamefont {Key}}, \bibinfo
  {author} {\bibfnamefont {S.~P.}\ \bibnamefont {Hatchett}}, \bibinfo {author}
  {\bibfnamefont {C.}~\bibnamefont {Brown}}, \bibinfo {author} {\bibfnamefont
  {W.}~\bibnamefont {Fountain}}, \bibinfo {author} {\bibfnamefont
  {J.}~\bibnamefont {Johnson}}, \bibinfo {author} {\bibfnamefont {D.~M.}\
  \bibnamefont {Pennington}}, \bibinfo {author} {\bibfnamefont {R.~A.}\
  \bibnamefont {Snavely}}, \bibinfo {author} {\bibfnamefont {S.~C.}\
  \bibnamefont {Wilks}}, \bibinfo {author} {\bibfnamefont {K.}~\bibnamefont
  {Yasuike}}, \bibinfo {author} {\bibfnamefont {H.}~\bibnamefont {Ruhl}},
  \bibinfo {author} {\bibfnamefont {F.}~\bibnamefont {Pegoraro}}, \bibinfo
  {author} {\bibfnamefont {S.~V.}\ \bibnamefont {Bulanov}}, \bibinfo {author}
  {\bibfnamefont {E.~M.}\ \bibnamefont {Campbell}}, \bibinfo {author}
  {\bibfnamefont {M.~D.}\ \bibnamefont {Perry}}, \ and\ \bibinfo {author}
  {\bibfnamefont {H.}~\bibnamefont {Powell}},\ }\bibfield  {title} {\enquote
  {\bibinfo {title} {Fast ignition by intense laser-accelerated proton
  beams},}\ }\href {\doibase 10.1103/PhysRevLett.86.436} {\bibfield  {journal}
  {\bibinfo  {journal} {Phys. Rev. Lett.}\ }\textbf {\bibinfo {volume} {86}},\
  \bibinfo {pages} {436--439} (\bibinfo {year} {2001})}\BibitemShut {NoStop}%
\bibitem [{\citenamefont {Naumova}\ \emph {et~al.}(2009)\citenamefont
  {Naumova}, \citenamefont {Schlegel}, \citenamefont {Tikhonchuk},
  \citenamefont {Labaune}, \citenamefont {Sokolov},\ and\ \citenamefont
  {Mourou}}]{Naumova_2009}%
  \BibitemOpen
  \bibfield  {author} {\bibinfo {author} {\bibfnamefont {N.}~\bibnamefont
  {Naumova}}, \bibinfo {author} {\bibfnamefont {T.}~\bibnamefont {Schlegel}},
  \bibinfo {author} {\bibfnamefont {V.~T.}\ \bibnamefont {Tikhonchuk}},
  \bibinfo {author} {\bibfnamefont {C.}~\bibnamefont {Labaune}}, \bibinfo
  {author} {\bibfnamefont {I.~V.}\ \bibnamefont {Sokolov}}, \ and\ \bibinfo
  {author} {\bibfnamefont {G.}~\bibnamefont {Mourou}},\ }\bibfield  {title}
  {\enquote {\bibinfo {title} {Hole boring in a {DT} pellet and fast-ion
  ignition with ultraintense laser pulses},}\ }\href {\doibase
  10.1103/PhysRevLett.102.025002} {\bibfield  {journal} {\bibinfo  {journal}
  {Phys. Rev. Lett.}\ }\textbf {\bibinfo {volume} {102}},\ \bibinfo {pages}
  {025002} (\bibinfo {year} {2009})}\BibitemShut {NoStop}%
\bibitem [{\citenamefont {Tikhonchuk}\ \emph {et~al.}(2010)\citenamefont
  {Tikhonchuk}, \citenamefont {Schlegel}, \citenamefont {Regan}, \citenamefont
  {Temporal}, \citenamefont {Feugeas}, \citenamefont {Nicolaï},\ and\
  \citenamefont {Ribeyre}}]{Tikhonchuk_2010}%
  \BibitemOpen
  \bibfield  {author} {\bibinfo {author} {\bibfnamefont {V.T.}\ \bibnamefont
  {Tikhonchuk}}, \bibinfo {author} {\bibfnamefont {T.}~\bibnamefont
  {Schlegel}}, \bibinfo {author} {\bibfnamefont {C.}~\bibnamefont {Regan}},
  \bibinfo {author} {\bibfnamefont {M.}~\bibnamefont {Temporal}}, \bibinfo
  {author} {\bibfnamefont {J.-L.}\ \bibnamefont {Feugeas}}, \bibinfo {author}
  {\bibfnamefont {Ph.}\ \bibnamefont {Nicolaï}}, \ and\ \bibinfo {author}
  {\bibfnamefont {X.}~\bibnamefont {Ribeyre}},\ }\bibfield  {title} {\enquote
  {\bibinfo {title} {Fast ion ignition with ultra-intense laser pulses},}\
  }\href {\doibase 10.1088/0029-5515/50/4/045003} {\bibfield  {journal}
  {\bibinfo  {journal} {Nucl. Fusion}\ }\textbf {\bibinfo {volume} {50}},\
  \bibinfo {pages} {045003} (\bibinfo {year} {2010})}\BibitemShut {NoStop}%
\bibitem [{\citenamefont {Macchi}\ \emph
  {et~al.}(2013{\natexlab{b}})\citenamefont {Macchi}, \citenamefont {Sgattoni},
  \citenamefont {Sinigardi}, \citenamefont {Borghesi},\ and\ \citenamefont
  {Passoni}}]{Macchi2013}%
  \BibitemOpen
  \bibfield  {author} {\bibinfo {author} {\bibfnamefont {A}~\bibnamefont
  {Macchi}}, \bibinfo {author} {\bibfnamefont {A}~\bibnamefont {Sgattoni}},
  \bibinfo {author} {\bibfnamefont {S}~\bibnamefont {Sinigardi}}, \bibinfo
  {author} {\bibfnamefont {M}~\bibnamefont {Borghesi}}, \ and\ \bibinfo
  {author} {\bibfnamefont {M}~\bibnamefont {Passoni}},\ }\bibfield  {title}
  {\enquote {\bibinfo {title} {Advanced strategies for ion acceleration using
  high-power lasers},}\ }\href {\doibase 10.1088/0741-3335/55/12/124020}
  {\bibfield  {journal} {\bibinfo  {journal} {Plasma Phys. Control. Fusion}\
  }\textbf {\bibinfo {volume} {55}},\ \bibinfo {pages} {124020} (\bibinfo
  {year} {2013}{\natexlab{b}})}\BibitemShut {NoStop}%
\bibitem [{LHC()}]{LHC}%
  \BibitemOpen
  \href@noop {} {\enquote {\bibinfo {title} {{CERN, Large Hadron Colliders
  (LHC)}, https://home.cern /science /accelerators /large-hadron-collider},}\
  }\BibitemShut {NoStop}%
\bibitem [{RHI()}]{RHIC}%
  \BibitemOpen
  \href@noop {} {\enquote {\bibinfo {title} {{RHIC (Relativistic Heavy Ion
  Colliders)}, https://www.bnl.gov /rhic/},}\ }\BibitemShut {NoStop}%
\bibitem [{\citenamefont {Yoon}\ \emph {et~al.}(2019)\citenamefont {Yoon},
  \citenamefont {Jeon}, \citenamefont {Shin}, \citenamefont {Lee},
  \citenamefont {Lee}, \citenamefont {Choi}, \citenamefont {Kim}, \citenamefont
  {Sung},\ and\ \citenamefont {Nam}}]{Yoon_2019_Achieving}%
  \BibitemOpen
  \bibfield  {author} {\bibinfo {author} {\bibfnamefont {Jin~Woo}\ \bibnamefont
  {Yoon}}, \bibinfo {author} {\bibfnamefont {Cheonha}\ \bibnamefont {Jeon}},
  \bibinfo {author} {\bibfnamefont {Junghoon}\ \bibnamefont {Shin}}, \bibinfo
  {author} {\bibfnamefont {Seong~Ku}\ \bibnamefont {Lee}}, \bibinfo {author}
  {\bibfnamefont {Hwang~Woon}\ \bibnamefont {Lee}}, \bibinfo {author}
  {\bibfnamefont {Il~Woo}\ \bibnamefont {Choi}}, \bibinfo {author}
  {\bibfnamefont {Hyung~Taek}\ \bibnamefont {Kim}}, \bibinfo {author}
  {\bibfnamefont {Jae~Hee}\ \bibnamefont {Sung}}, \ and\ \bibinfo {author}
  {\bibfnamefont {Chang~Hee}\ \bibnamefont {Nam}},\ }\bibfield  {title}
  {\enquote {\bibinfo {title} {{Achieving the laser intensity of
  $5.5\times10^{22} {\rm W/cm}^2$ with a wavefront-corrected multi-{PW}
  laser}},}\ }\href {\doibase 10.1364/oe.27.020412} {\bibfield  {journal}
  {\bibinfo  {journal} {Opt. Express}\ }\textbf {\bibinfo {volume} {27}},\
  \bibinfo {pages} {20412} (\bibinfo {year} {2019})}\BibitemShut {NoStop}%
\bibitem [{\citenamefont {Danson}\ \emph {et~al.}(2019)\citenamefont {Danson},
  \citenamefont {Haefner}, \citenamefont {Bromage}, \citenamefont {Butcher},
  \citenamefont {Chanteloup}, \citenamefont {Chowdhury}, \citenamefont
  {Galvanauskas}, \citenamefont {Gizzi}, \citenamefont {Hein}, \citenamefont
  {Hillier}, \citenamefont {Hopps}, \citenamefont {Kato}, \citenamefont
  {Khazanov}, \citenamefont {Kodama}, \citenamefont {Korn}, \citenamefont {Li},
  \citenamefont {Li}, \citenamefont {Limpert}, \citenamefont {Ma},
  \citenamefont {Nam}, \citenamefont {Neely}, \citenamefont {Papadopoulos},
  \citenamefont {Penman}, \citenamefont {Qian}, \citenamefont {Rocca},
  \citenamefont {Shaykin}, \citenamefont {Siders}, \citenamefont {Spindloe},
  \citenamefont {Szatm{\'{a}}ri}, \citenamefont {Trines}, \citenamefont {Zhu},
  \citenamefont {Zhu},\ and\ \citenamefont {Zuegel}}]{Danson_2019_Petawatt}%
  \BibitemOpen
  \bibfield  {author} {\bibinfo {author} {\bibfnamefont {Colin~N.}\
  \bibnamefont {Danson}}, \bibinfo {author} {\bibfnamefont {Constantin}\
  \bibnamefont {Haefner}}, \bibinfo {author} {\bibfnamefont {Jake}\
  \bibnamefont {Bromage}}, \bibinfo {author} {\bibfnamefont {Thomas}\
  \bibnamefont {Butcher}}, \bibinfo {author} {\bibfnamefont
  {Jean-Christophe~F.}\ \bibnamefont {Chanteloup}}, \bibinfo {author}
  {\bibfnamefont {Enam~A.}\ \bibnamefont {Chowdhury}}, \bibinfo {author}
  {\bibfnamefont {Almantas}\ \bibnamefont {Galvanauskas}}, \bibinfo {author}
  {\bibfnamefont {Leonida~A.}\ \bibnamefont {Gizzi}}, \bibinfo {author}
  {\bibfnamefont {Joachim}\ \bibnamefont {Hein}}, \bibinfo {author}
  {\bibfnamefont {David~I.}\ \bibnamefont {Hillier}}, \bibinfo {author}
  {\bibfnamefont {Nicholas~W.}\ \bibnamefont {Hopps}}, \bibinfo {author}
  {\bibfnamefont {Yoshiaki}\ \bibnamefont {Kato}}, \bibinfo {author}
  {\bibfnamefont {Efim~A.}\ \bibnamefont {Khazanov}}, \bibinfo {author}
  {\bibfnamefont {Ryosuke}\ \bibnamefont {Kodama}}, \bibinfo {author}
  {\bibfnamefont {Georg}\ \bibnamefont {Korn}}, \bibinfo {author}
  {\bibfnamefont {Ruxin}\ \bibnamefont {Li}}, \bibinfo {author} {\bibfnamefont
  {Yutong}\ \bibnamefont {Li}}, \bibinfo {author} {\bibfnamefont {Jens}\
  \bibnamefont {Limpert}}, \bibinfo {author} {\bibfnamefont {Jingui}\
  \bibnamefont {Ma}}, \bibinfo {author} {\bibfnamefont {Chang~Hee}\
  \bibnamefont {Nam}}, \bibinfo {author} {\bibfnamefont {David}\ \bibnamefont
  {Neely}}, \bibinfo {author} {\bibfnamefont {Dimitrios}\ \bibnamefont
  {Papadopoulos}}, \bibinfo {author} {\bibfnamefont {Rory~R.}\ \bibnamefont
  {Penman}}, \bibinfo {author} {\bibfnamefont {Liejia}\ \bibnamefont {Qian}},
  \bibinfo {author} {\bibfnamefont {Jorge~J.}\ \bibnamefont {Rocca}}, \bibinfo
  {author} {\bibfnamefont {Andrey~A.}\ \bibnamefont {Shaykin}}, \bibinfo
  {author} {\bibfnamefont {Craig~W.}\ \bibnamefont {Siders}}, \bibinfo {author}
  {\bibfnamefont {Christopher}\ \bibnamefont {Spindloe}}, \bibinfo {author}
  {\bibfnamefont {S{\'{a}}ndor}\ \bibnamefont {Szatm{\'{a}}ri}}, \bibinfo
  {author} {\bibfnamefont {Raoul M. G.~M.}\ \bibnamefont {Trines}}, \bibinfo
  {author} {\bibfnamefont {Jianqiang}\ \bibnamefont {Zhu}}, \bibinfo {author}
  {\bibfnamefont {Ping}\ \bibnamefont {Zhu}}, \ and\ \bibinfo {author}
  {\bibfnamefont {Jonathan~D.}\ \bibnamefont {Zuegel}},\ }\bibfield  {title}
  {\enquote {\bibinfo {title} {Petawatt and exawatt class lasers worldwide},}\
  }\href {\doibase 10.1017/hpl.2019.36} {\bibfield  {journal} {\bibinfo
  {journal} {High Power Laser Sci. Eng.}\ }\textbf {\bibinfo {volume} {7}},\
  \bibinfo {pages} {e54} (\bibinfo {year} {2019})}\BibitemShut {NoStop}%
\bibitem [{\citenamefont {Gales}\ \emph {et~al.}(2018)\citenamefont {Gales},
  \citenamefont {Tanaka}, \citenamefont {Balabanski}, \citenamefont {Negoita},
  \citenamefont {Stutman}, \citenamefont {Tesileanu}, \citenamefont {Ur},
  \citenamefont {Ursescu}, \citenamefont {Andrei}, \citenamefont {Ataman},
  \citenamefont {Cernaianu}, \citenamefont {D'Alessi}, \citenamefont {Dancus},
  \citenamefont {Diaconescu}, \citenamefont {Djourelov}, \citenamefont
  {Filipescu}, \citenamefont {Ghenuche}, \citenamefont {Ghita}, \citenamefont
  {Matei}, \citenamefont {Seto}, \citenamefont {Zeng},\ and\ \citenamefont
  {Zamfir}}]{Gales_2018_extreme}%
  \BibitemOpen
  \bibfield  {author} {\bibinfo {author} {\bibfnamefont {S}~\bibnamefont
  {Gales}}, \bibinfo {author} {\bibfnamefont {K~A}\ \bibnamefont {Tanaka}},
  \bibinfo {author} {\bibfnamefont {D~L}\ \bibnamefont {Balabanski}}, \bibinfo
  {author} {\bibfnamefont {F}~\bibnamefont {Negoita}}, \bibinfo {author}
  {\bibfnamefont {D}~\bibnamefont {Stutman}}, \bibinfo {author} {\bibfnamefont
  {O}~\bibnamefont {Tesileanu}}, \bibinfo {author} {\bibfnamefont {C~A}\
  \bibnamefont {Ur}}, \bibinfo {author} {\bibfnamefont {D}~\bibnamefont
  {Ursescu}}, \bibinfo {author} {\bibfnamefont {I}~\bibnamefont {Andrei}},
  \bibinfo {author} {\bibfnamefont {S}~\bibnamefont {Ataman}}, \bibinfo
  {author} {\bibfnamefont {M~O}\ \bibnamefont {Cernaianu}}, \bibinfo {author}
  {\bibfnamefont {L}~\bibnamefont {D'Alessi}}, \bibinfo {author} {\bibfnamefont
  {I}~\bibnamefont {Dancus}}, \bibinfo {author} {\bibfnamefont {B}~\bibnamefont
  {Diaconescu}}, \bibinfo {author} {\bibfnamefont {N}~\bibnamefont
  {Djourelov}}, \bibinfo {author} {\bibfnamefont {D}~\bibnamefont {Filipescu}},
  \bibinfo {author} {\bibfnamefont {P}~\bibnamefont {Ghenuche}}, \bibinfo
  {author} {\bibfnamefont {D~G}\ \bibnamefont {Ghita}}, \bibinfo {author}
  {\bibfnamefont {C}~\bibnamefont {Matei}}, \bibinfo {author} {\bibfnamefont
  {K}~\bibnamefont {Seto}}, \bibinfo {author} {\bibfnamefont {M}~\bibnamefont
  {Zeng}}, \ and\ \bibinfo {author} {\bibfnamefont {N~V}\ \bibnamefont
  {Zamfir}},\ }\bibfield  {title} {\enquote {\bibinfo {title} {The extreme
  light infrastructure{\textemdash}nuclear physics ({ELI}-{NP}) facility: new
  horizons in physics with 10 {PW} ultra-intense lasers and 20 {MeV} brilliant
  gamma beams},}\ }\href {\doibase 10.1088/1361-6633/aacfe8} {\bibfield
  {journal} {\bibinfo  {journal} {Rep. Progr. Phys.}\ }\textbf {\bibinfo
  {volume} {81}},\ \bibinfo {pages} {094301} (\bibinfo {year}
  {2018})}\BibitemShut {NoStop}%
\bibitem [{ELI()}]{ELI}%
  \BibitemOpen
  \href {www.extremelight-infrastructure.eu} {\enquote {\bibinfo {title} {{The
  Extreme Light Infrastructure (ELI)}},}\ }\bibinfo {howpublished}
  {http://www.eli-beams.eu/en/facility/lasers/}\BibitemShut {NoStop}%
\bibitem [{\citenamefont {Li}()}]{SULF}%
  \BibitemOpen
  \bibfield  {author} {\bibinfo {author} {\bibfnamefont {Ru~Xin}\ \bibnamefont
  {Li}},\ }\bibfield  {title} {\enquote {\bibinfo {title} {Progress of the
  {SULF 10 PW} laser project},}\ }in\ \href@noop {} {\emph {\bibinfo
  {booktitle} {1st AAPPS-DPP Meeting}}}\BibitemShut {NoStop}%
\bibitem [{Apo()}]{Apollon}%
  \BibitemOpen
  \href {www.polytechnique.edu} {\enquote {\bibinfo {title} {{Apollon multi-PW
  laser Users Facility}},}\ }\bibinfo {howpublished}
  {http://www.polytechnique.edu}\BibitemShut {NoStop}%
\bibitem [{\citenamefont {Higginson}\ \emph {et~al.}(2018)\citenamefont
  {Higginson}, \citenamefont {Gray}, \citenamefont {King}, \citenamefont
  {Dance}, \citenamefont {Williamson}, \citenamefont {Butler}, \citenamefont
  {Wilson}, \citenamefont {Capdessus}, \citenamefont {Armstrong}, \citenamefont
  {Green}, \citenamefont {Hawkes}, \citenamefont {Martin}, \citenamefont {Wei},
  \citenamefont {Mirfayzi}, \citenamefont {Yuan}, \citenamefont {Kar},
  \citenamefont {Borghesi}, \citenamefont {Clarke}, \citenamefont {Neely},\
  and\ \citenamefont {McKenna}}]{Higginson2018}%
  \BibitemOpen
  \bibfield  {author} {\bibinfo {author} {\bibfnamefont {A.}~\bibnamefont
  {Higginson}}, \bibinfo {author} {\bibfnamefont {R.~J.}\ \bibnamefont {Gray}},
  \bibinfo {author} {\bibfnamefont {M.}~\bibnamefont {King}}, \bibinfo {author}
  {\bibfnamefont {R.~J.}\ \bibnamefont {Dance}}, \bibinfo {author}
  {\bibfnamefont {S.~D.~R.}\ \bibnamefont {Williamson}}, \bibinfo {author}
  {\bibfnamefont {N.~M.~H.}\ \bibnamefont {Butler}}, \bibinfo {author}
  {\bibfnamefont {R.}~\bibnamefont {Wilson}}, \bibinfo {author} {\bibfnamefont
  {R.}~\bibnamefont {Capdessus}}, \bibinfo {author} {\bibfnamefont
  {C.}~\bibnamefont {Armstrong}}, \bibinfo {author} {\bibfnamefont {J.~S.}\
  \bibnamefont {Green}}, \bibinfo {author} {\bibfnamefont {S.~J.}\ \bibnamefont
  {Hawkes}}, \bibinfo {author} {\bibfnamefont {P.}~\bibnamefont {Martin}},
  \bibinfo {author} {\bibfnamefont {W.~Q.}\ \bibnamefont {Wei}}, \bibinfo
  {author} {\bibfnamefont {S.~R.}\ \bibnamefont {Mirfayzi}}, \bibinfo {author}
  {\bibfnamefont {X.~H.}\ \bibnamefont {Yuan}}, \bibinfo {author}
  {\bibfnamefont {S.}~\bibnamefont {Kar}}, \bibinfo {author} {\bibfnamefont
  {M.}~\bibnamefont {Borghesi}}, \bibinfo {author} {\bibfnamefont {R.~J.}\
  \bibnamefont {Clarke}}, \bibinfo {author} {\bibfnamefont {D.}~\bibnamefont
  {Neely}}, \ and\ \bibinfo {author} {\bibfnamefont {P.}~\bibnamefont
  {McKenna}},\ }\bibfield  {title} {\enquote {\bibinfo {title} {Near-100 {MeV}
  protons via a laser-driven transparency-enhanced hybrid acceleration
  scheme},}\ }\href {\doibase 10.1038/s41467-018-03063-9} {\bibfield  {journal}
  {\bibinfo  {journal} {Nat. Commun.}\ }\textbf {\bibinfo {volume} {9}},\
  \bibinfo {pages} {724} (\bibinfo {year} {2018})}\BibitemShut {NoStop}%
\bibitem [{\citenamefont {Haberberger}\ \emph {et~al.}(2011)\citenamefont
  {Haberberger}, \citenamefont {Tochitsky}, \citenamefont {Fiuza},
  \citenamefont {Gong}, \citenamefont {Fonseca}, \citenamefont {Silva},
  \citenamefont {Mori},\ and\ \citenamefont {Joshi}}]{Haberberger2011}%
  \BibitemOpen
  \bibfield  {author} {\bibinfo {author} {\bibfnamefont {Dan}\ \bibnamefont
  {Haberberger}}, \bibinfo {author} {\bibfnamefont {Sergei}\ \bibnamefont
  {Tochitsky}}, \bibinfo {author} {\bibfnamefont {Frederico}\ \bibnamefont
  {Fiuza}}, \bibinfo {author} {\bibfnamefont {Chao}\ \bibnamefont {Gong}},
  \bibinfo {author} {\bibfnamefont {Ricardo~A.}\ \bibnamefont {Fonseca}},
  \bibinfo {author} {\bibfnamefont {Luis~O.}\ \bibnamefont {Silva}}, \bibinfo
  {author} {\bibfnamefont {Warren~B.}\ \bibnamefont {Mori}}, \ and\ \bibinfo
  {author} {\bibfnamefont {Chan}\ \bibnamefont {Joshi}},\ }\bibfield  {title}
  {\enquote {\bibinfo {title} {Collisionless shocks in laser-produced plasma
  generate monoenergetic high-energy proton beams},}\ }\href {\doibase
  10.1038/nphys2130} {\bibfield  {journal} {\bibinfo  {journal} {Nat. Phys.}\
  }\textbf {\bibinfo {volume} {8}},\ \bibinfo {pages} {95--99} (\bibinfo {year}
  {2011})}\BibitemShut {NoStop}%
\bibitem [{\citenamefont {Zhang}\ \emph {et~al.}(2017)\citenamefont {Zhang},
  \citenamefont {Shen}, \citenamefont {Wang}, \citenamefont {Zhai},
  \citenamefont {Li}, \citenamefont {Lu}, \citenamefont {Li}, \citenamefont
  {Xu}, \citenamefont {Wang}, \citenamefont {Liang}, \citenamefont {Leng},
  \citenamefont {Li},\ and\ \citenamefont {Xu}}]{Zhang2017}%
  \BibitemOpen
  \bibfield  {author} {\bibinfo {author} {\bibfnamefont {H.}~\bibnamefont
  {Zhang}}, \bibinfo {author} {\bibfnamefont {B.{\hspace{0.167em}}F.}\
  \bibnamefont {Shen}}, \bibinfo {author} {\bibfnamefont
  {W.{\hspace{0.167em}}P.}\ \bibnamefont {Wang}}, \bibinfo {author}
  {\bibfnamefont {S.{\hspace{0.167em}}H.}\ \bibnamefont {Zhai}}, \bibinfo
  {author} {\bibfnamefont {S.{\hspace{0.167em}}S.}\ \bibnamefont {Li}},
  \bibinfo {author} {\bibfnamefont {X.{\hspace{0.167em}}M.}\ \bibnamefont
  {Lu}}, \bibinfo {author} {\bibfnamefont {J.{\hspace{0.167em}}F.}\
  \bibnamefont {Li}}, \bibinfo {author} {\bibfnamefont
  {R.{\hspace{0.167em}}J.}\ \bibnamefont {Xu}}, \bibinfo {author}
  {\bibfnamefont {X.{\hspace{0.167em}}L.}\ \bibnamefont {Wang}}, \bibinfo
  {author} {\bibfnamefont {X.{\hspace{0.167em}}Y.}\ \bibnamefont {Liang}},
  \bibinfo {author} {\bibfnamefont {Y.{\hspace{0.167em}}X.}\ \bibnamefont
  {Leng}}, \bibinfo {author} {\bibfnamefont {R.{\hspace{0.167em}}X.}\
  \bibnamefont {Li}}, \ and\ \bibinfo {author} {\bibfnamefont
  {Z.{\hspace{0.167em}}Z.}\ \bibnamefont {Xu}},\ }\bibfield  {title} {\enquote
  {\bibinfo {title} {Collisionless shock acceleration of high-flux
  quasimonoenergetic proton beams driven by circularly polarized laser
  pulses},}\ }\href {\doibase 10.1103/physrevlett.119.164801} {\bibfield
  {journal} {\bibinfo  {journal} {Phys. Rev. Lett.}\ }\textbf {\bibinfo
  {volume} {119}},\ \bibinfo {pages} {164801} (\bibinfo {year}
  {2017})}\BibitemShut {NoStop}%
\bibitem [{\citenamefont {Henig}\ \emph {et~al.}(2009)\citenamefont {Henig},
  \citenamefont {Steinke}, \citenamefont {Schnürer}, \citenamefont {Sokollik},
  \citenamefont {Hörlein}, \citenamefont {Kiefer}, \citenamefont {Jung},
  \citenamefont {Schreiber}, \citenamefont {Hegelich}, \citenamefont {Yan},
  \citenamefont {ter Vehn}, \citenamefont {Tajima}, \citenamefont {Nickles},
  \citenamefont {Sandner},\ and\ \citenamefont {Habs}}]{Henig2009}%
  \BibitemOpen
  \bibfield  {author} {\bibinfo {author} {\bibfnamefont {A.}~\bibnamefont
  {Henig}}, \bibinfo {author} {\bibfnamefont {S.}~\bibnamefont {Steinke}},
  \bibinfo {author} {\bibfnamefont {M.}~\bibnamefont {Schnürer}}, \bibinfo
  {author} {\bibfnamefont {T.}~\bibnamefont {Sokollik}}, \bibinfo {author}
  {\bibfnamefont {R.}~\bibnamefont {Hörlein}}, \bibinfo {author}
  {\bibfnamefont {D.}~\bibnamefont {Kiefer}}, \bibinfo {author} {\bibfnamefont
  {D.}~\bibnamefont {Jung}}, \bibinfo {author} {\bibfnamefont {J.}~\bibnamefont
  {Schreiber}}, \bibinfo {author} {\bibfnamefont {B.~M.}\ \bibnamefont
  {Hegelich}}, \bibinfo {author} {\bibfnamefont {X.~Q.}\ \bibnamefont {Yan}},
  \bibinfo {author} {\bibfnamefont {J.~Meyer}\ \bibnamefont {ter Vehn}},
  \bibinfo {author} {\bibfnamefont {T.}~\bibnamefont {Tajima}}, \bibinfo
  {author} {\bibfnamefont {P.~V.}\ \bibnamefont {Nickles}}, \bibinfo {author}
  {\bibfnamefont {W.}~\bibnamefont {Sandner}}, \ and\ \bibinfo {author}
  {\bibfnamefont {D.}~\bibnamefont {Habs}},\ }\bibfield  {title} {\enquote
  {\bibinfo {title} {Radiation-pressure acceleration of ion beams driven by
  circularly polarized laser pulses},}\ }\href {\doibase
  10.1103/physrevlett.103.245003} {\bibfield  {journal} {\bibinfo  {journal}
  {Phys. Rev. Lett.}\ }\textbf {\bibinfo {volume} {103}},\ \bibinfo {pages}
  {245003} (\bibinfo {year} {2009})}\BibitemShut {NoStop}%
\bibitem [{\citenamefont {Kar}\ \emph {et~al.}(2012)\citenamefont {Kar},
  \citenamefont {Kakolee}, \citenamefont {Qiao}, \citenamefont {Macchi},
  \citenamefont {Cerchez}, \citenamefont {Doria}, \citenamefont {Geissler},
  \citenamefont {McKenna}, \citenamefont {Neely}, \citenamefont {Osterholz},
  \citenamefont {Prasad}, \citenamefont {Quinn}, \citenamefont {Ramakrishna},
  \citenamefont {Sarri}, \citenamefont {Willi}, \citenamefont {Yuan},
  \citenamefont {Zepf},\ and\ \citenamefont {Borghesi}}]{Kar2012}%
  \BibitemOpen
  \bibfield  {author} {\bibinfo {author} {\bibfnamefont {S.}~\bibnamefont
  {Kar}}, \bibinfo {author} {\bibfnamefont {K.~F.}\ \bibnamefont {Kakolee}},
  \bibinfo {author} {\bibfnamefont {B.}~\bibnamefont {Qiao}}, \bibinfo {author}
  {\bibfnamefont {A.}~\bibnamefont {Macchi}}, \bibinfo {author} {\bibfnamefont
  {M.}~\bibnamefont {Cerchez}}, \bibinfo {author} {\bibfnamefont
  {D.}~\bibnamefont {Doria}}, \bibinfo {author} {\bibfnamefont
  {M.}~\bibnamefont {Geissler}}, \bibinfo {author} {\bibfnamefont
  {P.}~\bibnamefont {McKenna}}, \bibinfo {author} {\bibfnamefont
  {D.}~\bibnamefont {Neely}}, \bibinfo {author} {\bibfnamefont
  {J.}~\bibnamefont {Osterholz}}, \bibinfo {author} {\bibfnamefont
  {R.}~\bibnamefont {Prasad}}, \bibinfo {author} {\bibfnamefont
  {K.}~\bibnamefont {Quinn}}, \bibinfo {author} {\bibfnamefont
  {B.}~\bibnamefont {Ramakrishna}}, \bibinfo {author} {\bibfnamefont
  {G.}~\bibnamefont {Sarri}}, \bibinfo {author} {\bibfnamefont
  {O.}~\bibnamefont {Willi}}, \bibinfo {author} {\bibfnamefont {X.~Y.}\
  \bibnamefont {Yuan}}, \bibinfo {author} {\bibfnamefont {M.}~\bibnamefont
  {Zepf}}, \ and\ \bibinfo {author} {\bibfnamefont {M.}~\bibnamefont
  {Borghesi}},\ }\bibfield  {title} {\enquote {\bibinfo {title} {Ion
  acceleration in multispecies targets driven by intense laser radiation
  pressure},}\ }\href {\doibase 10.1103/physrevlett.109.185006} {\bibfield
  {journal} {\bibinfo  {journal} {Phys. Rev. Lett.}\ }\textbf {\bibinfo
  {volume} {109}},\ \bibinfo {pages} {185006} (\bibinfo {year}
  {2012})}\BibitemShut {NoStop}%
\bibitem [{\citenamefont {Bin}\ \emph {et~al.}(2015)\citenamefont {Bin},
  \citenamefont {Ma}, \citenamefont {Wang}, \citenamefont {Streeter},
  \citenamefont {Kreuzer}, \citenamefont {Kiefer}, \citenamefont {Yeung},
  \citenamefont {Cousens}, \citenamefont {Foster}, \citenamefont {Dromey},
  \citenamefont {Yan}, \citenamefont {Ramis}, \citenamefont {ter Vehn},
  \citenamefont {Zepf},\ and\ \citenamefont {Schreiber}}]{Bin2015}%
  \BibitemOpen
  \bibfield  {author} {\bibinfo {author} {\bibfnamefont
  {J.{\hspace{0.167em}}H.}\ \bibnamefont {Bin}}, \bibinfo {author}
  {\bibfnamefont {W.{\hspace{0.167em}}J.}\ \bibnamefont {Ma}}, \bibinfo
  {author} {\bibfnamefont {H.{\hspace{0.167em}}Y.}\ \bibnamefont {Wang}},
  \bibinfo {author} {\bibfnamefont
  {M.{\hspace{0.167em}}J.{\hspace{0.167em}}V.}\ \bibnamefont {Streeter}},
  \bibinfo {author} {\bibfnamefont {C.}~\bibnamefont {Kreuzer}}, \bibinfo
  {author} {\bibfnamefont {D.}~\bibnamefont {Kiefer}}, \bibinfo {author}
  {\bibfnamefont {M.}~\bibnamefont {Yeung}}, \bibinfo {author} {\bibfnamefont
  {S.}~\bibnamefont {Cousens}}, \bibinfo {author} {\bibfnamefont
  {P.{\hspace{0.167em}}S.}\ \bibnamefont {Foster}}, \bibinfo {author}
  {\bibfnamefont {B.}~\bibnamefont {Dromey}}, \bibinfo {author} {\bibfnamefont
  {X.{\hspace{0.167em}}Q.}\ \bibnamefont {Yan}}, \bibinfo {author}
  {\bibfnamefont {R.}~\bibnamefont {Ramis}}, \bibinfo {author} {\bibfnamefont
  {J.~Meyer}\ \bibnamefont {ter Vehn}}, \bibinfo {author} {\bibfnamefont
  {M.}~\bibnamefont {Zepf}}, \ and\ \bibinfo {author} {\bibfnamefont
  {J.}~\bibnamefont {Schreiber}},\ }\bibfield  {title} {\enquote {\bibinfo
  {title} {Ion acceleration using relativistic pulse shaping in
  near-critical-density plasmas},}\ }\href {\doibase
  10.1103/physrevlett.115.064801} {\bibfield  {journal} {\bibinfo  {journal}
  {Phys. Rev. Lett.}\ }\textbf {\bibinfo {volume} {115}},\ \bibinfo {pages}
  {064801} (\bibinfo {year} {2015})}\BibitemShut {NoStop}%
\bibitem [{\citenamefont {Scullion}\ \emph {et~al.}(2017)\citenamefont
  {Scullion}, \citenamefont {Doria}, \citenamefont {Romagnani}, \citenamefont
  {Sgattoni}, \citenamefont {Naughton}, \citenamefont {Symes}, \citenamefont
  {McKenna}, \citenamefont {Macchi}, \citenamefont {Zepf}, \citenamefont
  {Kar},\ and\ \citenamefont {Borghesi}}]{Scullion2017}%
  \BibitemOpen
  \bibfield  {author} {\bibinfo {author} {\bibfnamefont {C.}~\bibnamefont
  {Scullion}}, \bibinfo {author} {\bibfnamefont {D.}~\bibnamefont {Doria}},
  \bibinfo {author} {\bibfnamefont {L.}~\bibnamefont {Romagnani}}, \bibinfo
  {author} {\bibfnamefont {A.}~\bibnamefont {Sgattoni}}, \bibinfo {author}
  {\bibfnamefont {K.}~\bibnamefont {Naughton}}, \bibinfo {author}
  {\bibfnamefont {D.{\hspace{0.167em}}R.}\ \bibnamefont {Symes}}, \bibinfo
  {author} {\bibfnamefont {P.}~\bibnamefont {McKenna}}, \bibinfo {author}
  {\bibfnamefont {A.}~\bibnamefont {Macchi}}, \bibinfo {author} {\bibfnamefont
  {M.}~\bibnamefont {Zepf}}, \bibinfo {author} {\bibfnamefont {S.}~\bibnamefont
  {Kar}}, \ and\ \bibinfo {author} {\bibfnamefont {M.}~\bibnamefont
  {Borghesi}},\ }\bibfield  {title} {\enquote {\bibinfo {title} {Polarization
  dependence of bulk ion acceleration from ultrathin foils irradiated by
  high-intensity ultrashort laser pulses},}\ }\href {\doibase
  10.1103/physrevlett.119.054801} {\bibfield  {journal} {\bibinfo  {journal}
  {Phys. Rev. Lett.}\ }\textbf {\bibinfo {volume} {119}},\ \bibinfo {pages}
  {054801} (\bibinfo {year} {2017})}\BibitemShut {NoStop}%
\bibitem [{\citenamefont {Esirkepov}\ \emph {et~al.}(2004)\citenamefont
  {Esirkepov}, \citenamefont {Borghesi}, \citenamefont {Bulanov}, \citenamefont
  {Mourou},\ and\ \citenamefont {Tajima}}]{Esirkepov_2004}%
  \BibitemOpen
  \bibfield  {author} {\bibinfo {author} {\bibfnamefont {T.}~\bibnamefont
  {Esirkepov}}, \bibinfo {author} {\bibfnamefont {M.}~\bibnamefont {Borghesi}},
  \bibinfo {author} {\bibfnamefont {S.~V.}\ \bibnamefont {Bulanov}}, \bibinfo
  {author} {\bibfnamefont {G.}~\bibnamefont {Mourou}}, \ and\ \bibinfo {author}
  {\bibfnamefont {T.}~\bibnamefont {Tajima}},\ }\bibfield  {title} {\enquote
  {\bibinfo {title} {Highly efficient relativistic-ion generation in the
  laser-piston regime},}\ }\href {\doibase 10.1103/physrevlett.92.175003}
  {\bibfield  {journal} {\bibinfo  {journal} {Phys. Rev. Lett.}\ }\textbf
  {\bibinfo {volume} {92}},\ \bibinfo {pages} {175003} (\bibinfo {year}
  {2004})}\BibitemShut {NoStop}%
\bibitem [{\citenamefont {Pegoraro}\ and\ \citenamefont
  {Bulanov}(2007)}]{Pegoraro_2007}%
  \BibitemOpen
  \bibfield  {author} {\bibinfo {author} {\bibfnamefont {F.}~\bibnamefont
  {Pegoraro}}\ and\ \bibinfo {author} {\bibfnamefont {S.~V.}\ \bibnamefont
  {Bulanov}},\ }\bibfield  {title} {\enquote {\bibinfo {title} {Photon bubbles
  and ion acceleration in a plasma dominated by the radiation pressure of an
  electromagnetic pulse},}\ }\href {\doibase 10.1103/physrevlett.99.065002}
  {\bibfield  {journal} {\bibinfo  {journal} {Phys. Rev. Lett.}\ }\textbf
  {\bibinfo {volume} {99}},\ \bibinfo {pages} {065002} (\bibinfo {year}
  {2007})}\BibitemShut {NoStop}%
\bibitem [{\citenamefont {Macchi}\ \emph {et~al.}(2005)\citenamefont {Macchi},
  \citenamefont {Cattani}, \citenamefont {Liseykina},\ and\ \citenamefont
  {Cornolti}}]{Macchi2005}%
  \BibitemOpen
  \bibfield  {author} {\bibinfo {author} {\bibfnamefont {Andrea}\ \bibnamefont
  {Macchi}}, \bibinfo {author} {\bibfnamefont {Federica}\ \bibnamefont
  {Cattani}}, \bibinfo {author} {\bibfnamefont {Tatiana~V.}\ \bibnamefont
  {Liseykina}}, \ and\ \bibinfo {author} {\bibfnamefont {Fulvio}\ \bibnamefont
  {Cornolti}},\ }\bibfield  {title} {\enquote {\bibinfo {title} {Laser
  acceleration of ion bunches at the front surface of overdense plasmas},}\
  }\href {\doibase 10.1103/physrevlett.94.165003} {\bibfield  {journal}
  {\bibinfo  {journal} {Phys. Rev. Lett.}\ }\textbf {\bibinfo {volume} {94}},\
  \bibinfo {pages} {165003} (\bibinfo {year} {2005})}\BibitemShut {NoStop}%
\bibitem [{\citenamefont {Chen}\ \emph {et~al.}(2009)\citenamefont {Chen},
  \citenamefont {Pukhov}, \citenamefont {Yu},\ and\ \citenamefont
  {Sheng}}]{Chen2009}%
  \BibitemOpen
  \bibfield  {author} {\bibinfo {author} {\bibfnamefont {M.}~\bibnamefont
  {Chen}}, \bibinfo {author} {\bibfnamefont {A.}~\bibnamefont {Pukhov}},
  \bibinfo {author} {\bibfnamefont {T.~P.}\ \bibnamefont {Yu}}, \ and\ \bibinfo
  {author} {\bibfnamefont {Z.~M.}\ \bibnamefont {Sheng}},\ }\bibfield  {title}
  {\enquote {\bibinfo {title} {Enhanced collimated gev monoenergetic ion
  acceleration from a shaped foil target irradiated by a circularly polarized
  laser pulse},}\ }\href {\doibase 10.1103/PhysRevLett.103.024801} {\bibfield
  {journal} {\bibinfo  {journal} {Phys. Rev. Lett.}\ }\textbf {\bibinfo
  {volume} {103}},\ \bibinfo {pages} {024801} (\bibinfo {year}
  {2009})}\BibitemShut {NoStop}%
\bibitem [{\citenamefont {Yu}\ \emph {et~al.}(2010)\citenamefont {Yu},
  \citenamefont {Pukhov}, \citenamefont {Shvets},\ and\ \citenamefont
  {Chen}}]{Yu2010}%
  \BibitemOpen
  \bibfield  {author} {\bibinfo {author} {\bibfnamefont {Tong-Pu}\ \bibnamefont
  {Yu}}, \bibinfo {author} {\bibfnamefont {Alexander}\ \bibnamefont {Pukhov}},
  \bibinfo {author} {\bibfnamefont {Gennady}\ \bibnamefont {Shvets}}, \ and\
  \bibinfo {author} {\bibfnamefont {Min}\ \bibnamefont {Chen}},\ }\bibfield
  {title} {\enquote {\bibinfo {title} {Stable laser-driven proton beam
  acceleration from a two-ion-species ultrathin foil},}\ }\href {\doibase
  10.1103/PhysRevLett.105.065002} {\bibfield  {journal} {\bibinfo  {journal}
  {Phys. Rev. Lett.}\ }\textbf {\bibinfo {volume} {105}},\ \bibinfo {pages}
  {065002} (\bibinfo {year} {2010})}\BibitemShut {NoStop}%
\bibitem [{\citenamefont {Bulanov}\ \emph {et~al.}(2010)\citenamefont
  {Bulanov}, \citenamefont {Echkina}, \citenamefont {Esirkepov}, \citenamefont
  {Inovenkov}, \citenamefont {Kando}, \citenamefont {Pegoraro},\ and\
  \citenamefont {Korn}}]{Bulanov_2010}%
  \BibitemOpen
  \bibfield  {author} {\bibinfo {author} {\bibfnamefont {S.~V.}\ \bibnamefont
  {Bulanov}}, \bibinfo {author} {\bibfnamefont {E.~Yu.}\ \bibnamefont
  {Echkina}}, \bibinfo {author} {\bibfnamefont {T.~Zh.}\ \bibnamefont
  {Esirkepov}}, \bibinfo {author} {\bibfnamefont {I.~N.}\ \bibnamefont
  {Inovenkov}}, \bibinfo {author} {\bibfnamefont {M.}~\bibnamefont {Kando}},
  \bibinfo {author} {\bibfnamefont {F.}~\bibnamefont {Pegoraro}}, \ and\
  \bibinfo {author} {\bibfnamefont {G.}~\bibnamefont {Korn}},\ }\bibfield
  {title} {\enquote {\bibinfo {title} {Unlimited ion acceleration by radiation
  pressure},}\ }\href {\doibase 10.1103/PhysRevLett.104.135003} {\bibfield
  {journal} {\bibinfo  {journal} {Phys. Rev. Lett.}\ }\textbf {\bibinfo
  {volume} {104}},\ \bibinfo {pages} {135003} (\bibinfo {year}
  {2010})}\BibitemShut {NoStop}%
\bibitem [{\citenamefont {Ji}\ \emph {et~al.}(2014)\citenamefont {Ji},
  \citenamefont {Pukhov},\ and\ \citenamefont {Shen}}]{Ji_2014}%
  \BibitemOpen
  \bibfield  {author} {\bibinfo {author} {\bibfnamefont {Liangliang}\
  \bibnamefont {Ji}}, \bibinfo {author} {\bibfnamefont {Alexander}\
  \bibnamefont {Pukhov}}, \ and\ \bibinfo {author} {\bibfnamefont {Baifei}\
  \bibnamefont {Shen}},\ }\bibfield  {title} {\enquote {\bibinfo {title} {Ion
  acceleration in the `dragging field' of a light-pressure-driven piston},}\
  }\href {\doibase 10.1088/1367-2630/16/6/063047} {\bibfield  {journal}
  {\bibinfo  {journal} {New J. Phys.}\ }\textbf {\bibinfo {volume} {16}},\
  \bibinfo {pages} {063047} (\bibinfo {year} {2014})}\BibitemShut {NoStop}%
\bibitem [{\citenamefont {Zhou}\ \emph {et~al.}(2016)\citenamefont {Zhou},
  \citenamefont {Yan}, \citenamefont {Mourou}, \citenamefont {Wheeler},
  \citenamefont {Bin}, \citenamefont {Schreiber},\ and\ \citenamefont
  {Tajima}}]{Zhou2016}%
  \BibitemOpen
  \bibfield  {author} {\bibinfo {author} {\bibfnamefont {M.~L.}\ \bibnamefont
  {Zhou}}, \bibinfo {author} {\bibfnamefont {X.~Q.}\ \bibnamefont {Yan}},
  \bibinfo {author} {\bibfnamefont {G.}~\bibnamefont {Mourou}}, \bibinfo
  {author} {\bibfnamefont {J.~A.}\ \bibnamefont {Wheeler}}, \bibinfo {author}
  {\bibfnamefont {J.~H.}\ \bibnamefont {Bin}}, \bibinfo {author} {\bibfnamefont
  {J.}~\bibnamefont {Schreiber}}, \ and\ \bibinfo {author} {\bibfnamefont
  {T.}~\bibnamefont {Tajima}},\ }\bibfield  {title} {\enquote {\bibinfo {title}
  {Proton acceleration by single-cycle laser pulses offers a novel
  monoenergetic and stable operating regime},}\ }\href {\doibase
  10.1063/1.4947544} {\bibfield  {journal} {\bibinfo  {journal} {Phys.
  Plasmas}\ }\textbf {\bibinfo {volume} {23}},\ \bibinfo {pages} {043112}
  (\bibinfo {year} {2016})}\BibitemShut {NoStop}%
\bibitem [{\citenamefont {Wan}\ \emph {et~al.}(2020{\natexlab{a}})\citenamefont
  {Wan}, \citenamefont {Andriyash}, \citenamefont {Pai}, \citenamefont {Hua},
  \citenamefont {Zhang}, \citenamefont {Li}, \citenamefont {Wu}, \citenamefont
  {Nie}, \citenamefont {Mori}, \citenamefont {Lu}, \citenamefont {Malka},\ and\
  \citenamefont {Joshi}}]{Wan2020a}%
  \BibitemOpen
  \bibfield  {author} {\bibinfo {author} {\bibfnamefont {Y}~\bibnamefont
  {Wan}}, \bibinfo {author} {\bibfnamefont {I~A}\ \bibnamefont {Andriyash}},
  \bibinfo {author} {\bibfnamefont {C~H}\ \bibnamefont {Pai}}, \bibinfo
  {author} {\bibfnamefont {J~F}\ \bibnamefont {Hua}}, \bibinfo {author}
  {\bibfnamefont {C~J}\ \bibnamefont {Zhang}}, \bibinfo {author} {\bibfnamefont
  {F}~\bibnamefont {Li}}, \bibinfo {author} {\bibfnamefont {Y~P}\ \bibnamefont
  {Wu}}, \bibinfo {author} {\bibfnamefont {Z}~\bibnamefont {Nie}}, \bibinfo
  {author} {\bibfnamefont {W~B}\ \bibnamefont {Mori}}, \bibinfo {author}
  {\bibfnamefont {W}~\bibnamefont {Lu}}, \bibinfo {author} {\bibfnamefont
  {V}~\bibnamefont {Malka}}, \ and\ \bibinfo {author} {\bibfnamefont
  {C}~\bibnamefont {Joshi}},\ }\bibfield  {title} {\enquote {\bibinfo {title}
  {Ion acceleration with an ultra-intense two-frequency laser tweezer},}\
  }\href {\doibase 10.1088/1367-2630/ab8678} {\bibfield  {journal} {\bibinfo
  {journal} {New J. Phys.}\ }\textbf {\bibinfo {volume} {22}},\ \bibinfo
  {pages} {052002} (\bibinfo {year} {2020}{\natexlab{a}})}\BibitemShut
  {NoStop}%
\bibitem [{\citenamefont {Landau}\ and\ \citenamefont
  {Lifshitz}(1975)}]{Landau1975}%
  \BibitemOpen
  \bibfield  {author} {\bibinfo {author} {\bibfnamefont {L.~D.}\ \bibnamefont
  {Landau}}\ and\ \bibinfo {author} {\bibfnamefont {E.~M.}\ \bibnamefont
  {Lifshitz}},\ }\href@noop {} {\emph {\bibinfo {title} {The Classical Theory
  of Fields}}}\ (\bibinfo  {publisher} {Elsevier, Oxford},\ \bibinfo {year}
  {1975})\BibitemShut {NoStop}%
\bibitem [{\citenamefont {Tamburini}\ \emph {et~al.}(2010)\citenamefont
  {Tamburini}, \citenamefont {Pegoraro}, \citenamefont {Di~Piazza},
  \citenamefont {Keitel},\ and\ \citenamefont {Macchi}}]{Tamburini2010}%
  \BibitemOpen
  \bibfield  {author} {\bibinfo {author} {\bibfnamefont {M.}~\bibnamefont
  {Tamburini}}, \bibinfo {author} {\bibfnamefont {F.}~\bibnamefont {Pegoraro}},
  \bibinfo {author} {\bibfnamefont {A.}~\bibnamefont {Di~Piazza}}, \bibinfo
  {author} {\bibfnamefont {C.~H.}\ \bibnamefont {Keitel}}, \ and\ \bibinfo
  {author} {\bibfnamefont {A.}~\bibnamefont {Macchi}},\ }\bibfield  {title}
  {\enquote {\bibinfo {title} {Radiation reaction effects on radiation pressure
  acceleration},}\ }\href {\doibase 10.1088/1367-2630/12/12/123005} {\bibfield
  {journal} {\bibinfo  {journal} {New J. Phys.}\ }\textbf {\bibinfo {volume}
  {12}},\ \bibinfo {pages} {123005} (\bibinfo {year} {2010})}\BibitemShut
  {NoStop}%
\bibitem [{\citenamefont {Tamburini}\ \emph {et~al.}(2011)\citenamefont
  {Tamburini}, \citenamefont {Pegoraro}, \citenamefont {Di~Piazza},
  \citenamefont {Keitel}, \citenamefont {Liseykina},\ and\ \citenamefont
  {Macchi}}]{Tamburini2011}%
  \BibitemOpen
  \bibfield  {author} {\bibinfo {author} {\bibfnamefont {M.}~\bibnamefont
  {Tamburini}}, \bibinfo {author} {\bibfnamefont {F.}~\bibnamefont {Pegoraro}},
  \bibinfo {author} {\bibfnamefont {A.}~\bibnamefont {Di~Piazza}}, \bibinfo
  {author} {\bibfnamefont {C.~H.}\ \bibnamefont {Keitel}}, \bibinfo {author}
  {\bibfnamefont {T.~V.}\ \bibnamefont {Liseykina}}, \ and\ \bibinfo {author}
  {\bibfnamefont {A.}~\bibnamefont {Macchi}},\ }\bibfield  {title} {\enquote
  {\bibinfo {title} {Radiation reaction effects on electron nonlinear dynamics
  and ion acceleration in laser-solid interaction},}\ }\href@noop {} {\bibfield
   {journal} {\bibinfo  {journal} {Nucl. Instrum. Methods Phys. Res., Sect. A}\
  }\textbf {\bibinfo {volume} {653}},\ \bibinfo {pages} {181} (\bibinfo {year}
  {2011})}\BibitemShut {NoStop}%
\bibitem [{\citenamefont {Piazza}\ \emph {et~al.}(2012)\citenamefont {Piazza},
  \citenamefont {Müller}, \citenamefont {Hatsagortsyan},\ and\ \citenamefont
  {Keitel}}]{Piazza2012}%
  \BibitemOpen
  \bibfield  {author} {\bibinfo {author} {\bibfnamefont {A.~Di}\ \bibnamefont
  {Piazza}}, \bibinfo {author} {\bibfnamefont {C.}~\bibnamefont {Müller}},
  \bibinfo {author} {\bibfnamefont {K.~Z.}\ \bibnamefont {Hatsagortsyan}}, \
  and\ \bibinfo {author} {\bibfnamefont {C.~H.}\ \bibnamefont {Keitel}},\
  }\bibfield  {title} {\enquote {\bibinfo {title} {Extremely high-intensity
  laser interactions with fundamental quantum systems},}\ }\href {\doibase
  10.1103/revmodphys.84.1177} {\bibfield  {journal} {\bibinfo  {journal} {Rev.
  Mod. Phys.}\ }\textbf {\bibinfo {volume} {84}},\ \bibinfo {pages}
  {1177--1228} (\bibinfo {year} {2012})}\BibitemShut {NoStop}%
\bibitem [{\citenamefont {Chen}\ \emph {et~al.}(2010)\citenamefont {Chen},
  \citenamefont {Pukhov}, \citenamefont {Yu},\ and\ \citenamefont
  {Sheng}}]{Chen_2010}%
  \BibitemOpen
  \bibfield  {author} {\bibinfo {author} {\bibfnamefont {Min}\ \bibnamefont
  {Chen}}, \bibinfo {author} {\bibfnamefont {Alexander}\ \bibnamefont
  {Pukhov}}, \bibinfo {author} {\bibfnamefont {Tong-Pu}\ \bibnamefont {Yu}}, \
  and\ \bibinfo {author} {\bibfnamefont {Zheng-Ming}\ \bibnamefont {Sheng}},\
  }\bibfield  {title} {\enquote {\bibinfo {title} {Radiation reaction effects
  on ion acceleration in laser foil interaction},}\ }\href {\doibase
  10.1088/0741-3335/53/1/014004} {\bibfield  {journal} {\bibinfo  {journal}
  {Plasma Phys. Control. Fusion}\ }\textbf {\bibinfo {volume} {53}},\ \bibinfo
  {pages} {014004} (\bibinfo {year} {2010})}\BibitemShut {NoStop}%
\bibitem [{\citenamefont {Capdessus}\ and\ \citenamefont
  {McKenna}(2015)}]{Capdessus2015}%
  \BibitemOpen
  \bibfield  {author} {\bibinfo {author} {\bibfnamefont {R.}~\bibnamefont
  {Capdessus}}\ and\ \bibinfo {author} {\bibfnamefont {P.}~\bibnamefont
  {McKenna}},\ }\bibfield  {title} {\enquote {\bibinfo {title} {Influence of
  radiation reaction force on ultraintense laser-driven ion acceleration},}\
  }\href {\doibase 10.1103/PhysRevE.91.053105} {\bibfield  {journal} {\bibinfo
  {journal} {Phys. Rev. E}\ }\textbf {\bibinfo {volume} {91}},\ \bibinfo
  {pages} {053105} (\bibinfo {year} {2015})}\BibitemShut {NoStop}%
\bibitem [{\citenamefont {Wan}\ \emph {et~al.}(2019)\citenamefont {Wan},
  \citenamefont {Xue}, \citenamefont {Dou}, \citenamefont {Hatsagortsyan},
  \citenamefont {Yan}, \citenamefont {Khikhlukha}, \citenamefont {Bulanov},
  \citenamefont {Korn}, \citenamefont {Zhao}, \citenamefont {Xu},\ and\
  \citenamefont {Li}}]{Wan_2019}%
  \BibitemOpen
  \bibfield  {author} {\bibinfo {author} {\bibfnamefont {F.}~\bibnamefont
  {Wan}}, \bibinfo {author} {\bibfnamefont {K.}~\bibnamefont {Xue}}, \bibinfo
  {author} {\bibfnamefont {Z.-K.}\ \bibnamefont {Dou}}, \bibinfo {author}
  {\bibfnamefont {K.~Z.}\ \bibnamefont {Hatsagortsyan}}, \bibinfo {author}
  {\bibfnamefont {W.}~\bibnamefont {Yan}}, \bibinfo {author} {\bibfnamefont
  {D.}~\bibnamefont {Khikhlukha}}, \bibinfo {author} {\bibfnamefont {S.~V.}\
  \bibnamefont {Bulanov}}, \bibinfo {author} {\bibfnamefont {G.}~\bibnamefont
  {Korn}}, \bibinfo {author} {\bibfnamefont {Y.-T.}\ \bibnamefont {Zhao}},
  \bibinfo {author} {\bibfnamefont {Z.-F.}\ \bibnamefont {Xu}}, \ and\ \bibinfo
  {author} {\bibfnamefont {J.-X.}\ \bibnamefont {Li}},\ }\bibfield  {title}
  {\enquote {\bibinfo {title} {Imprint of the stochastic nature of photon
  emission by electrons on the proton energy spectra in the laser-plasma
  interaction},}\ }\href@noop {} {\bibfield  {journal} {\bibinfo  {journal}
  {Plasma Phys. Control. Fusion}\ }\textbf {\bibinfo {volume} {61}},\ \bibinfo
  {pages} {084010} (\bibinfo {year} {2019})}\BibitemShut {NoStop}%
\bibitem [{\citenamefont {Seipt}\ \emph {et~al.}(2018)\citenamefont {Seipt},
  \citenamefont {Sorbo}, \citenamefont {Ridgers},\ and\ \citenamefont
  {Thomas}}]{Seipt2018}%
  \BibitemOpen
  \bibfield  {author} {\bibinfo {author} {\bibfnamefont {D.}~\bibnamefont
  {Seipt}}, \bibinfo {author} {\bibfnamefont {D.~Del}\ \bibnamefont {Sorbo}},
  \bibinfo {author} {\bibfnamefont {C.~P.}\ \bibnamefont {Ridgers}}, \ and\
  \bibinfo {author} {\bibfnamefont {A.~G.~R.}\ \bibnamefont {Thomas}},\
  }\bibfield  {title} {\enquote {\bibinfo {title} {Theory of radiative electron
  polarization in strong laser fields},}\ }\href {<Go to
  ISI>://WOS:000442194400009
  https://journals.aps.org/pra/pdf/10.1103/PhysRevA.98.023417} {\bibfield
  {journal} {\bibinfo  {journal} {Phys. Rev. A}\ }\textbf {\bibinfo {volume}
  {98}},\ \bibinfo {pages} {023417} (\bibinfo {year} {2018})}\BibitemShut
  {NoStop}%
\bibitem [{\citenamefont {Büscher}\ \emph {et~al.}(2020)\citenamefont
  {Büscher}, \citenamefont {Hützen}, \citenamefont {Ji},\ and\ \citenamefont
  {Lehrach}}]{Buescher2020}%
  \BibitemOpen
  \bibfield  {author} {\bibinfo {author} {\bibfnamefont {Markus}\ \bibnamefont
  {Büscher}}, \bibinfo {author} {\bibfnamefont {Anna}\ \bibnamefont
  {Hützen}}, \bibinfo {author} {\bibfnamefont {Liangliang}\ \bibnamefont
  {Ji}}, \ and\ \bibinfo {author} {\bibfnamefont {Andreas}\ \bibnamefont
  {Lehrach}},\ }\bibfield  {title} {\enquote {\bibinfo {title} {Generation of
  polarized particle beams at relativistic laser intensities},}\ }\href
  {\doibase 10.1017/hpl.2020.35} {\bibfield  {journal} {\bibinfo  {journal}
  {High Power Laser Sci. Eng.}\ }\textbf {\bibinfo {volume} {8}},\ \bibinfo
  {pages} {e36} (\bibinfo {year} {2020})}\BibitemShut {NoStop}%
\bibitem [{\citenamefont {Xue}\ \emph {et~al.}(2020)\citenamefont {Xue},
  \citenamefont {Dou}, \citenamefont {Wan}, \citenamefont {Yu}, \citenamefont
  {Wang}, \citenamefont {Ren}, \citenamefont {Zhao}, \citenamefont {Zhao},
  \citenamefont {Xu},\ and\ \citenamefont {Li}}]{Xue2020}%
  \BibitemOpen
  \bibfield  {author} {\bibinfo {author} {\bibfnamefont {Kun}\ \bibnamefont
  {Xue}}, \bibinfo {author} {\bibfnamefont {Zhen-Ke}\ \bibnamefont {Dou}},
  \bibinfo {author} {\bibfnamefont {Feng}\ \bibnamefont {Wan}}, \bibinfo
  {author} {\bibfnamefont {Tong-Pu}\ \bibnamefont {Yu}}, \bibinfo {author}
  {\bibfnamefont {Wei-Min}\ \bibnamefont {Wang}}, \bibinfo {author}
  {\bibfnamefont {Jie-Ru}\ \bibnamefont {Ren}}, \bibinfo {author}
  {\bibfnamefont {Qian}\ \bibnamefont {Zhao}}, \bibinfo {author} {\bibfnamefont
  {Yong-Tao}\ \bibnamefont {Zhao}}, \bibinfo {author} {\bibfnamefont
  {Zhong-Feng}\ \bibnamefont {Xu}}, \ and\ \bibinfo {author} {\bibfnamefont
  {Jian-Xing}\ \bibnamefont {Li}},\ }\bibfield  {title} {\enquote {\bibinfo
  {title} {Generation of highly-polarized high-energy brilliant $\gamma$-rays
  via laser-plasma interaction},}\ }\href {\doibase 10.1063/5.0007734}
  {\bibfield  {journal} {\bibinfo  {journal} {Matter Radiat. Extremes}\
  }\textbf {\bibinfo {volume} {5}},\ \bibinfo {pages} {054402} (\bibinfo {year}
  {2020})}\BibitemShut {NoStop}%
\bibitem [{\citenamefont {Li}\ \emph {et~al.}(2019)\citenamefont {Li},
  \citenamefont {Shaisultanov}, \citenamefont {Hatsagortsyan}, \citenamefont
  {Wan}, \citenamefont {Keitel},\ and\ \citenamefont {Li}}]{li2019prl}%
  \BibitemOpen
  \bibfield  {author} {\bibinfo {author} {\bibfnamefont {Yan-Fei}\ \bibnamefont
  {Li}}, \bibinfo {author} {\bibfnamefont {Rashid}\ \bibnamefont
  {Shaisultanov}}, \bibinfo {author} {\bibfnamefont {Karen~Z.}\ \bibnamefont
  {Hatsagortsyan}}, \bibinfo {author} {\bibfnamefont {Feng}\ \bibnamefont
  {Wan}}, \bibinfo {author} {\bibfnamefont {Christoph~H.}\ \bibnamefont
  {Keitel}}, \ and\ \bibinfo {author} {\bibfnamefont {Jian-Xing}\ \bibnamefont
  {Li}},\ }\bibfield  {title} {\enquote {\bibinfo {title} {Ultrarelativistic
  electron-beam polarization in single-shot interaction with an ultraintense
  laser pulse},}\ }\href {\doibase 10.1103/PhysRevLett.122.154801} {\bibfield
  {journal} {\bibinfo  {journal} {Phys. Rev. Lett.}\ }\textbf {\bibinfo
  {volume} {122}},\ \bibinfo {pages} {154801} (\bibinfo {year}
  {2019})}\BibitemShut {NoStop}%
\bibitem [{\citenamefont {Li}\ \emph {et~al.}(2020)\citenamefont {Li},
  \citenamefont {Shaisultanov}, \citenamefont {Chen}, \citenamefont {Wan},
  \citenamefont {Hatsagortsyan}, \citenamefont {Keitel},\ and\ \citenamefont
  {Li}}]{Ligammaray_2020}%
  \BibitemOpen
  \bibfield  {author} {\bibinfo {author} {\bibfnamefont {Yan-Fei}\ \bibnamefont
  {Li}}, \bibinfo {author} {\bibfnamefont {Rashid}\ \bibnamefont
  {Shaisultanov}}, \bibinfo {author} {\bibfnamefont {Yue-Yue}\ \bibnamefont
  {Chen}}, \bibinfo {author} {\bibfnamefont {Feng}\ \bibnamefont {Wan}},
  \bibinfo {author} {\bibfnamefont {Karen~Z.}\ \bibnamefont {Hatsagortsyan}},
  \bibinfo {author} {\bibfnamefont {Christoph~H.}\ \bibnamefont {Keitel}}, \
  and\ \bibinfo {author} {\bibfnamefont {Jian-Xing}\ \bibnamefont {Li}},\
  }\bibfield  {title} {\enquote {\bibinfo {title} {Polarized ultrashort
  brilliant multi-gev $\ensuremath{\gamma}$ rays via single-shot laser-electron
  interaction},}\ }\href {\doibase 10.1103/PhysRevLett.124.014801} {\bibfield
  {journal} {\bibinfo  {journal} {Phys. Rev. Lett.}\ }\textbf {\bibinfo
  {volume} {124}},\ \bibinfo {pages} {014801} (\bibinfo {year}
  {2020})}\BibitemShut {NoStop}%
\bibitem [{\citenamefont {Liu}\ \emph {et~al.}(2020)\citenamefont {Liu},
  \citenamefont {Xue}, \citenamefont {Wan}, \citenamefont {Chen}, \citenamefont
  {Li}, \citenamefont {Liu}, \citenamefont {Weng}, \citenamefont {Sheng},\ and\
  \citenamefont {Zhang}}]{Liu2020}%
  \BibitemOpen
  \bibfield  {author} {\bibinfo {author} {\bibfnamefont {Wei-Yuan}\
  \bibnamefont {Liu}}, \bibinfo {author} {\bibfnamefont {Kun}\ \bibnamefont
  {Xue}}, \bibinfo {author} {\bibfnamefont {Feng}\ \bibnamefont {Wan}},
  \bibinfo {author} {\bibfnamefont {Min}\ \bibnamefont {Chen}}, \bibinfo
  {author} {\bibfnamefont {Jian-Xing}\ \bibnamefont {Li}}, \bibinfo {author}
  {\bibfnamefont {Feng}\ \bibnamefont {Liu}}, \bibinfo {author} {\bibfnamefont
  {Su-Ming}\ \bibnamefont {Weng}}, \bibinfo {author} {\bibfnamefont
  {Zheng-Ming}\ \bibnamefont {Sheng}}, \ and\ \bibinfo {author} {\bibfnamefont
  {Jie}\ \bibnamefont {Zhang}},\ }\bibfield  {title} {\enquote {\bibinfo
  {title} {Trapping and acceleration of spin-polarized positrons from $\gamma$
  photon splitting in wakefields},}\ }\href@noop {} {\bibfield  {journal}
  {\bibinfo  {journal} {Arxiv}\ } (\bibinfo {year} {2020})},\ \Eprint
  {http://arxiv.org/abs/2011.00156} {arXiv:2011.00156 [physics.plasm-ph]}
  \BibitemShut {NoStop}%
\bibitem [{\citenamefont {Guo}\ \emph {et~al.}(2020)\citenamefont {Guo},
  \citenamefont {Wang}, \citenamefont {Shaisultanov}, \citenamefont {Wan},
  \citenamefont {Xu}, \citenamefont {Chen}, \citenamefont {Hatsagortsyan},\
  and\ \citenamefont {Li}}]{Guo_2020}%
  \BibitemOpen
  \bibfield  {author} {\bibinfo {author} {\bibfnamefont {Ren-Tong}\
  \bibnamefont {Guo}}, \bibinfo {author} {\bibfnamefont {Yu}~\bibnamefont
  {Wang}}, \bibinfo {author} {\bibfnamefont {Rashid}\ \bibnamefont
  {Shaisultanov}}, \bibinfo {author} {\bibfnamefont {Feng}\ \bibnamefont
  {Wan}}, \bibinfo {author} {\bibfnamefont {Zhong-Feng}\ \bibnamefont {Xu}},
  \bibinfo {author} {\bibfnamefont {Yue-Yue}\ \bibnamefont {Chen}}, \bibinfo
  {author} {\bibfnamefont {Karen~Z.}\ \bibnamefont {Hatsagortsyan}}, \ and\
  \bibinfo {author} {\bibfnamefont {Jian-Xing}\ \bibnamefont {Li}},\ }\bibfield
   {title} {\enquote {\bibinfo {title} {Stochasticity in radiative polarization
  of ultrarelativistic electrons in an ultrastrong laser pulse},}\ }\href
  {\doibase 10.1103/PhysRevResearch.2.033483} {\bibfield  {journal} {\bibinfo
  {journal} {Phys. Rev. Research}\ }\textbf {\bibinfo {volume} {2}},\ \bibinfo
  {pages} {033483} (\bibinfo {year} {2020})}\BibitemShut {NoStop}%
\bibitem [{\citenamefont {Arber}\ \emph {et~al.}(2015)\citenamefont {Arber},
  \citenamefont {Bennett}, \citenamefont {Brady}, \citenamefont
  {Lawrence-Douglas}, \citenamefont {Ramsay}, \citenamefont {Sircombe},
  \citenamefont {Gillies}, \citenamefont {Evans}, \citenamefont {Schmitz},
  \citenamefont {Bell},\ and\ \citenamefont {Ridgers}}]{Arber2015}%
  \BibitemOpen
  \bibfield  {author} {\bibinfo {author} {\bibfnamefont {T~D}\ \bibnamefont
  {Arber}}, \bibinfo {author} {\bibfnamefont {K}~\bibnamefont {Bennett}},
  \bibinfo {author} {\bibfnamefont {C~S}\ \bibnamefont {Brady}}, \bibinfo
  {author} {\bibfnamefont {A}~\bibnamefont {Lawrence-Douglas}}, \bibinfo
  {author} {\bibfnamefont {M~G}\ \bibnamefont {Ramsay}}, \bibinfo {author}
  {\bibfnamefont {N~J}\ \bibnamefont {Sircombe}}, \bibinfo {author}
  {\bibfnamefont {P}~\bibnamefont {Gillies}}, \bibinfo {author} {\bibfnamefont
  {R~G}\ \bibnamefont {Evans}}, \bibinfo {author} {\bibfnamefont
  {H}~\bibnamefont {Schmitz}}, \bibinfo {author} {\bibfnamefont {A~R}\
  \bibnamefont {Bell}}, \ and\ \bibinfo {author} {\bibfnamefont {C~P}\
  \bibnamefont {Ridgers}},\ }\bibfield  {title} {\enquote {\bibinfo {title}
  {Contemporary particle-in-cell approach to laser-plasma modelling},}\ }\href
  {\doibase 10.1088/0741-3335/57/11/113001} {\bibfield  {journal} {\bibinfo
  {journal} {Plasma Phys. Control. Fusion}\ }\textbf {\bibinfo {volume} {57}},\
  \bibinfo {pages} {113001} (\bibinfo {year} {2015})}\BibitemShut {NoStop}%
\bibitem [{\citenamefont {Macchi}\ \emph {et~al.}(2009)\citenamefont {Macchi},
  \citenamefont {Veghini},\ and\ \citenamefont {Pegoraro}}]{Macchi2009}%
  \BibitemOpen
  \bibfield  {author} {\bibinfo {author} {\bibfnamefont {Andrea}\ \bibnamefont
  {Macchi}}, \bibinfo {author} {\bibfnamefont {Silvia}\ \bibnamefont
  {Veghini}}, \ and\ \bibinfo {author} {\bibfnamefont {Francesco}\ \bibnamefont
  {Pegoraro}},\ }\bibfield  {title} {\enquote {\bibinfo {title}
  {{\textquotedblleft}light sail{\textquotedblright} acceleration
  reexamined},}\ }\href {\doibase 10.1103/physrevlett.103.085003} {\bibfield
  {journal} {\bibinfo  {journal} {Phys. Rev. Lett.}\ }\textbf {\bibinfo
  {volume} {103}},\ \bibinfo {pages} {085003} (\bibinfo {year}
  {2009})}\BibitemShut {NoStop}%
\bibitem [{\citenamefont {Qiao}\ \emph {et~al.}(2009)\citenamefont {Qiao},
  \citenamefont {Zepf}, \citenamefont {Borghesi},\ and\ \citenamefont
  {Geissler}}]{Qiao2009}%
  \BibitemOpen
  \bibfield  {author} {\bibinfo {author} {\bibfnamefont {B.}~\bibnamefont
  {Qiao}}, \bibinfo {author} {\bibfnamefont {M.}~\bibnamefont {Zepf}}, \bibinfo
  {author} {\bibfnamefont {M.}~\bibnamefont {Borghesi}}, \ and\ \bibinfo
  {author} {\bibfnamefont {M.}~\bibnamefont {Geissler}},\ }\bibfield  {title}
  {\enquote {\bibinfo {title} {Stable {GeV} ion-beam acceleration from thin
  foils by circularly polarized laser pulses},}\ }\href {\doibase
  10.1103/PhysRevLett.102.145002} {\bibfield  {journal} {\bibinfo  {journal}
  {Phys. Rev. Lett.}\ }\textbf {\bibinfo {volume} {102}},\ \bibinfo {pages}
  {145002} (\bibinfo {year} {2009})}\BibitemShut {NoStop}%
\bibitem [{\citenamefont {Qiao}\ \emph {et~al.}(2010)\citenamefont {Qiao},
  \citenamefont {Zepf}, \citenamefont {Borghesi}, \citenamefont {Dromey},
  \citenamefont {Geissler}, \citenamefont {Karmakar},\ and\ \citenamefont
  {Gibbon}}]{Qiao2010}%
  \BibitemOpen
  \bibfield  {author} {\bibinfo {author} {\bibfnamefont {B.}~\bibnamefont
  {Qiao}}, \bibinfo {author} {\bibfnamefont {M.}~\bibnamefont {Zepf}}, \bibinfo
  {author} {\bibfnamefont {M.}~\bibnamefont {Borghesi}}, \bibinfo {author}
  {\bibfnamefont {B.}~\bibnamefont {Dromey}}, \bibinfo {author} {\bibfnamefont
  {M.}~\bibnamefont {Geissler}}, \bibinfo {author} {\bibfnamefont
  {A.}~\bibnamefont {Karmakar}}, \ and\ \bibinfo {author} {\bibfnamefont
  {P.}~\bibnamefont {Gibbon}},\ }\bibfield  {title} {\enquote {\bibinfo {title}
  {Radiation-pressure acceleration of ion beams from nanofoil targets: The
  leaky light-sail regime},}\ }\href {\doibase 10.1103/PhysRevLett.105.155002}
  {\bibfield  {journal} {\bibinfo  {journal} {Phys. Rev. Lett.}\ }\textbf
  {\bibinfo {volume} {105}},\ \bibinfo {pages} {155002} (\bibinfo {year}
  {2010})}\BibitemShut {NoStop}%
\bibitem [{sup()}]{supplemental}%
  \BibitemOpen
  \href@noop {} {}\bibinfo {note} {See supplemental materials for details on
  analytical expressions of relativistic ponderomotive force and relative
  slippage distance, electron energy spectrum, and phase space evolution of
  protons.}\BibitemShut {Stop}%
\bibitem [{\citenamefont {Wan}\ \emph {et~al.}(2020{\natexlab{b}})\citenamefont
  {Wan}, \citenamefont {Wang}, \citenamefont {Guo}, \citenamefont {Chen},
  \citenamefont {Shaisultanov}, \citenamefont {Xu}, \citenamefont
  {Hatsagortsyan}, \citenamefont {Keitel},\ and\ \citenamefont
  {Li}}]{Wan_2020}%
  \BibitemOpen
  \bibfield  {author} {\bibinfo {author} {\bibfnamefont {Feng}\ \bibnamefont
  {Wan}}, \bibinfo {author} {\bibfnamefont {Yu}~\bibnamefont {Wang}}, \bibinfo
  {author} {\bibfnamefont {Ren-Tong}\ \bibnamefont {Guo}}, \bibinfo {author}
  {\bibfnamefont {Yue-Yue}\ \bibnamefont {Chen}}, \bibinfo {author}
  {\bibfnamefont {Rashid}\ \bibnamefont {Shaisultanov}}, \bibinfo {author}
  {\bibfnamefont {Zhong-Feng}\ \bibnamefont {Xu}}, \bibinfo {author}
  {\bibfnamefont {Karen~Z.}\ \bibnamefont {Hatsagortsyan}}, \bibinfo {author}
  {\bibfnamefont {Christoph~H.}\ \bibnamefont {Keitel}}, \ and\ \bibinfo
  {author} {\bibfnamefont {Jian-Xing}\ \bibnamefont {Li}},\ }\bibfield  {title}
  {\enquote {\bibinfo {title} {High-energy $\ensuremath{\gamma}$-photon
  polarization in nonlinear breit-wheeler pair production and
  $\ensuremath{\gamma}$ polarimetry},}\ }\href {\doibase
  10.1103/PhysRevResearch.2.032049} {\bibfield  {journal} {\bibinfo  {journal}
  {Phys. Rev. Research}\ }\textbf {\bibinfo {volume} {2}},\ \bibinfo {pages}
  {032049} (\bibinfo {year} {2020}{\natexlab{b}})}\BibitemShut {NoStop}%
\bibitem [{\citenamefont {Thomas}(1926)}]{Thomas_1926}%
  \BibitemOpen
  \bibfield  {author} {\bibinfo {author} {\bibfnamefont {L.~H.}\ \bibnamefont
  {Thomas}},\ }\bibfield  {title} {\enquote {\bibinfo {title} {The motion of
  the spinning electron},}\ }\href@noop {} {\bibfield  {journal} {\bibinfo
  {journal} {Nature (London)}\ }\textbf {\bibinfo {volume} {117}},\ \bibinfo
  {pages} {514} (\bibinfo {year} {1926})}\BibitemShut {NoStop}%
\bibitem [{\citenamefont {Thomas}(1927)}]{Thomas_1927}%
  \BibitemOpen
  \bibfield  {author} {\bibinfo {author} {\bibfnamefont {L.~H.}\ \bibnamefont
  {Thomas}},\ }\bibfield  {title} {\enquote {\bibinfo {title} {The kinematics
  of an electron with an axis},}\ }\href@noop {} {\bibfield  {journal}
  {\bibinfo  {journal} {Philos. Mag.}\ }\textbf {\bibinfo {volume} {3}},\
  \bibinfo {pages} {1--22} (\bibinfo {year} {1927})}\BibitemShut {NoStop}%
\bibitem [{\citenamefont {Bargmann}\ \emph {et~al.}(1959)\citenamefont
  {Bargmann}, \citenamefont {Michel},\ and\ \citenamefont
  {Telegdi}}]{Bargmann_1959}%
  \BibitemOpen
  \bibfield  {author} {\bibinfo {author} {\bibfnamefont {V.}~\bibnamefont
  {Bargmann}}, \bibinfo {author} {\bibfnamefont {Louis}\ \bibnamefont
  {Michel}}, \ and\ \bibinfo {author} {\bibfnamefont {V.~L.}\ \bibnamefont
  {Telegdi}},\ }\bibfield  {title} {\enquote {\bibinfo {title} {Precession of
  the polarization of particles moving in a homogeneous electromagnetic
  field},}\ }\href {\doibase 10.1103/PhysRevLett.2.435} {\bibfield  {journal}
  {\bibinfo  {journal} {Phys. Rev. Lett.}\ }\textbf {\bibinfo {volume} {2}},\
  \bibinfo {pages} {435--436} (\bibinfo {year} {1959})}\BibitemShut {NoStop}%
\bibitem [{\citenamefont {King}\ \emph {et~al.}(1999)\citenamefont {King},
  \citenamefont {Ables}, \citenamefont {Adams}, \citenamefont {Alrick},
  \citenamefont {Amann}, \citenamefont {Balzar}, \citenamefont {Jr},
  \citenamefont {Crow}, \citenamefont {Cushing}, \citenamefont {Eddleman},
  \citenamefont {Fife}, \citenamefont {Flores}, \citenamefont {Fujino},
  \citenamefont {Gallegos}, \citenamefont {Gray}, \citenamefont {Hartouni},
  \citenamefont {Hogan}, \citenamefont {Holmes}, \citenamefont {Jaramillo},
  \citenamefont {Knudsson}, \citenamefont {London}, \citenamefont {Lopez},
  \citenamefont {McDonald}, \citenamefont {McClelland}, \citenamefont
  {Merrill}, \citenamefont {Morley}, \citenamefont {Morris}, \citenamefont
  {Naivar}, \citenamefont {Parker}, \citenamefont {Park}, \citenamefont
  {Pazuchanics}, \citenamefont {Pillai}, \citenamefont {Riedel}, \citenamefont
  {Sarracino}, \citenamefont {Jr}, \citenamefont {Stacy}, \citenamefont
  {Takala}, \citenamefont {Thompson}, \citenamefont {Tucker}, \citenamefont
  {Yates}, \citenamefont {Ziock},\ and\ \citenamefont {Zumbro}}]{King1999}%
  \BibitemOpen
  \bibfield  {author} {\bibinfo {author} {\bibfnamefont {N.S.P}\ \bibnamefont
  {King}}, \bibinfo {author} {\bibfnamefont {E}~\bibnamefont {Ables}}, \bibinfo
  {author} {\bibfnamefont {Ken}\ \bibnamefont {Adams}}, \bibinfo {author}
  {\bibfnamefont {K.R}\ \bibnamefont {Alrick}}, \bibinfo {author}
  {\bibfnamefont {J.F}\ \bibnamefont {Amann}}, \bibinfo {author} {\bibfnamefont
  {Stephen}\ \bibnamefont {Balzar}}, \bibinfo {author} {\bibfnamefont
  {P.D~Barnes}\ \bibnamefont {Jr}}, \bibinfo {author} {\bibfnamefont {M.L}\
  \bibnamefont {Crow}}, \bibinfo {author} {\bibfnamefont {S.B}\ \bibnamefont
  {Cushing}}, \bibinfo {author} {\bibfnamefont {J.C}\ \bibnamefont {Eddleman}},
  \bibinfo {author} {\bibfnamefont {T.T}\ \bibnamefont {Fife}}, \bibinfo
  {author} {\bibfnamefont {Paul}\ \bibnamefont {Flores}}, \bibinfo {author}
  {\bibfnamefont {D}~\bibnamefont {Fujino}}, \bibinfo {author} {\bibfnamefont
  {R.A}\ \bibnamefont {Gallegos}}, \bibinfo {author} {\bibfnamefont {N.T}\
  \bibnamefont {Gray}}, \bibinfo {author} {\bibfnamefont {E.P}\ \bibnamefont
  {Hartouni}}, \bibinfo {author} {\bibfnamefont {G.E}\ \bibnamefont {Hogan}},
  \bibinfo {author} {\bibfnamefont {V.H}\ \bibnamefont {Holmes}}, \bibinfo
  {author} {\bibfnamefont {S.A}\ \bibnamefont {Jaramillo}}, \bibinfo {author}
  {\bibfnamefont {J.N}\ \bibnamefont {Knudsson}}, \bibinfo {author}
  {\bibfnamefont {R.K}\ \bibnamefont {London}}, \bibinfo {author}
  {\bibfnamefont {R.R}\ \bibnamefont {Lopez}}, \bibinfo {author} {\bibfnamefont
  {T.E}\ \bibnamefont {McDonald}}, \bibinfo {author} {\bibfnamefont {J.B}\
  \bibnamefont {McClelland}}, \bibinfo {author} {\bibfnamefont {F.E}\
  \bibnamefont {Merrill}}, \bibinfo {author} {\bibfnamefont {K.B}\ \bibnamefont
  {Morley}}, \bibinfo {author} {\bibfnamefont {C.L}\ \bibnamefont {Morris}},
  \bibinfo {author} {\bibfnamefont {F.J}\ \bibnamefont {Naivar}}, \bibinfo
  {author} {\bibfnamefont {E.L}\ \bibnamefont {Parker}}, \bibinfo {author}
  {\bibfnamefont {H.S}\ \bibnamefont {Park}}, \bibinfo {author} {\bibfnamefont
  {P.D}\ \bibnamefont {Pazuchanics}}, \bibinfo {author} {\bibfnamefont
  {C}~\bibnamefont {Pillai}}, \bibinfo {author} {\bibfnamefont {C.M}\
  \bibnamefont {Riedel}}, \bibinfo {author} {\bibfnamefont {J.S}\ \bibnamefont
  {Sarracino}}, \bibinfo {author} {\bibfnamefont {F.E~Shelley}\ \bibnamefont
  {Jr}}, \bibinfo {author} {\bibfnamefont {H.L}\ \bibnamefont {Stacy}},
  \bibinfo {author} {\bibfnamefont {B.E}\ \bibnamefont {Takala}}, \bibinfo
  {author} {\bibfnamefont {Richard}\ \bibnamefont {Thompson}}, \bibinfo
  {author} {\bibfnamefont {H.E}\ \bibnamefont {Tucker}}, \bibinfo {author}
  {\bibfnamefont {G.J}\ \bibnamefont {Yates}}, \bibinfo {author} {\bibfnamefont
  {H.-J}\ \bibnamefont {Ziock}}, \ and\ \bibinfo {author} {\bibfnamefont {J.D}\
  \bibnamefont {Zumbro}},\ }\bibfield  {title} {\enquote {\bibinfo {title} {An
  800-{MeV} proton radiography facility for dynamic experiments},}\ }\href
  {\doibase 10.1016/s0168-9002(98)01241-8} {\bibfield  {journal} {\bibinfo
  {journal} {Nucl. Instrum. Meth. A.}\ }\textbf {\bibinfo {volume} {424}},\
  \bibinfo {pages} {84--91} (\bibinfo {year} {1999})}\BibitemShut {NoStop}%
\bibitem [{\citenamefont {Wang}\ \emph {et~al.}(2020)\citenamefont {Wang},
  \citenamefont {Jiang}, \citenamefont {Dong}, \citenamefont {Lu},
  \citenamefont {Li}, \citenamefont {Xu}, \citenamefont {Sun}, \citenamefont
  {Yu}, \citenamefont {Guo}, \citenamefont {Liang}, \citenamefont {Leng},
  \citenamefont {Li},\ and\ \citenamefont {Xu}}]{Wang2020}%
  \BibitemOpen
  \bibfield  {author} {\bibinfo {author} {\bibfnamefont
  {W.{\hspace{0.167em}}P.}\ \bibnamefont {Wang}}, \bibinfo {author}
  {\bibfnamefont {C.}~\bibnamefont {Jiang}}, \bibinfo {author} {\bibfnamefont
  {H.}~\bibnamefont {Dong}}, \bibinfo {author} {\bibfnamefont
  {X.{\hspace{0.167em}}M.}\ \bibnamefont {Lu}}, \bibinfo {author}
  {\bibfnamefont {J.{\hspace{0.167em}}F.}\ \bibnamefont {Li}}, \bibinfo
  {author} {\bibfnamefont {R.{\hspace{0.167em}}J.}\ \bibnamefont {Xu}},
  \bibinfo {author} {\bibfnamefont {Y.{\hspace{0.167em}}J.}\ \bibnamefont
  {Sun}}, \bibinfo {author} {\bibfnamefont {L.{\hspace{0.167em}}H.}\
  \bibnamefont {Yu}}, \bibinfo {author} {\bibfnamefont {Z.}~\bibnamefont
  {Guo}}, \bibinfo {author} {\bibfnamefont {X.{\hspace{0.167em}}Y.}\
  \bibnamefont {Liang}}, \bibinfo {author} {\bibfnamefont
  {Y.{\hspace{0.167em}}X.}\ \bibnamefont {Leng}}, \bibinfo {author}
  {\bibfnamefont {R.{\hspace{0.167em}}X.}\ \bibnamefont {Li}}, \ and\ \bibinfo
  {author} {\bibfnamefont {Z.{\hspace{0.167em}}Z.}\ \bibnamefont {Xu}},\
  }\bibfield  {title} {\enquote {\bibinfo {title} {Hollow plasma acceleration
  driven by a relativistic reflected hollow laser},}\ }\href {\doibase
  10.1103/physrevlett.125.034801} {\bibfield  {journal} {\bibinfo  {journal}
  {Phys. Rev. Lett.}\ }\textbf {\bibinfo {volume} {125}},\ \bibinfo {pages}
  {034801} (\bibinfo {year} {2020})}\BibitemShut {NoStop}%
\bibitem [{\citenamefont {Avetissian}(2006)}]{Avetissian2006}%
  \BibitemOpen
  \bibfield  {author} {\bibinfo {author} {\bibfnamefont {Hamlet~Karo}\
  \bibnamefont {Avetissian}},\ }\href@noop {} {\emph {\bibinfo {title}
  {Relativistic nonlinear electrodynamics: interaction of charged particles
  with strong and super strong laser fields}}},\ Springer series in optical
  sciences,\ (\bibinfo  {publisher} {Springer},\ \bibinfo {address} {New
  York},\ \bibinfo {year} {2006})\ pp.\ \bibinfo {pages} {xiii, 333
  p.}\BibitemShut {Stop}%
\bibitem [{\citenamefont {Lindman}\ and\ \citenamefont
  {Stroscio}(1977)}]{Lindman1977}%
  \BibitemOpen
  \bibfield  {author} {\bibinfo {author} {\bibfnamefont {E.L.}\ \bibnamefont
  {Lindman}}\ and\ \bibinfo {author} {\bibfnamefont {M.A.}\ \bibnamefont
  {Stroscio}},\ }\bibfield  {title} {\enquote {\bibinfo {title} {On the
  relativistic corrections to the ponderomotive force},}\ }\href {\doibase
  10.1088/0029-5515/17/3/019} {\bibfield  {journal} {\bibinfo  {journal} {Nucl.
  Fusion}\ }\textbf {\bibinfo {volume} {17}},\ \bibinfo {pages} {619--621}
  (\bibinfo {year} {1977})}\BibitemShut {NoStop}%
\bibitem [{\citenamefont {Quesnel}\ and\ \citenamefont
  {Mora}(1998)}]{Quesnel_1998}%
  \BibitemOpen
  \bibfield  {author} {\bibinfo {author} {\bibfnamefont {Brice}\ \bibnamefont
  {Quesnel}}\ and\ \bibinfo {author} {\bibfnamefont {Patrick}\ \bibnamefont
  {Mora}},\ }\bibfield  {title} {\enquote {\bibinfo {title} {Theory and
  simulation of the interaction of ultraintense laser pulses with electrons in
  vacuum},}\ }\href@noop {} {\bibfield  {journal} {\bibinfo  {journal} {Phys.
  Rev. E}\ }\textbf {\bibinfo {volume} {58}},\ \bibinfo {pages} {3719--3732}
  (\bibinfo {year} {1998})}\BibitemShut {NoStop}%
\bibitem [{\citenamefont {Niel}\ \emph {et~al.}(2018)\citenamefont {Niel},
  \citenamefont {Riconda}, \citenamefont {Amiranoff}, \citenamefont {Duclous},\
  and\ \citenamefont {Grech}}]{Niel2018}%
  \BibitemOpen
  \bibfield  {author} {\bibinfo {author} {\bibfnamefont {F.}~\bibnamefont
  {Niel}}, \bibinfo {author} {\bibfnamefont {C.}~\bibnamefont {Riconda}},
  \bibinfo {author} {\bibfnamefont {F.}~\bibnamefont {Amiranoff}}, \bibinfo
  {author} {\bibfnamefont {R.}~\bibnamefont {Duclous}}, \ and\ \bibinfo
  {author} {\bibfnamefont {M.}~\bibnamefont {Grech}},\ }\bibfield  {title}
  {\enquote {\bibinfo {title} {From quantum to classical modeling of radiation
  reaction: A focus on stochasticity effects},}\ }\href {\doibase
  10.1103/PhysRevE.97.043209} {\bibfield  {journal} {\bibinfo  {journal} {Phys.
  Rev. E}\ }\textbf {\bibinfo {volume} {97}},\ \bibinfo {pages} {043209}
  (\bibinfo {year} {2018})}\BibitemShut {NoStop}%
\bibitem [{\citenamefont {Baier}\ \emph {et~al.}(1998)\citenamefont {Baier},
  \citenamefont {Katkov},\ and\ \citenamefont {Strakhovenko}}]{Baier1998}%
  \BibitemOpen
  \bibfield  {author} {\bibinfo {author} {\bibfnamefont {V.~N.}\ \bibnamefont
  {Baier}}, \bibinfo {author} {\bibfnamefont {V.~M.}\ \bibnamefont {Katkov}}, \
  and\ \bibinfo {author} {\bibfnamefont {V.~M.}\ \bibnamefont {Strakhovenko}},\
  }\href@noop {} {\emph {\bibinfo {title} {Electromagnetic Processes at High
  Energies in Oriented Single Crystals}}}\ (\bibinfo  {publisher} {World
  Scientific},\ \bibinfo {year} {1998})\BibitemShut {NoStop}%
\bibitem [{\citenamefont {Bulanov}\ \emph {et~al.}(2013)\citenamefont
  {Bulanov}, \citenamefont {Esirkepov}, \citenamefont {Kando}, \citenamefont
  {Koga}, \citenamefont {Nakamura}, \citenamefont {Bulanov}, \citenamefont
  {Zhidkov}, \citenamefont {Kato},\ and\ \citenamefont {Korn}}]{Bulanov2013a}%
  \BibitemOpen
  \bibfield  {author} {\bibinfo {author} {\bibfnamefont {Sergei~V.}\
  \bibnamefont {Bulanov}}, \bibinfo {author} {\bibfnamefont {Timur~Zh.}\
  \bibnamefont {Esirkepov}}, \bibinfo {author} {\bibfnamefont {Masaki}\
  \bibnamefont {Kando}}, \bibinfo {author} {\bibfnamefont {James~K.}\
  \bibnamefont {Koga}}, \bibinfo {author} {\bibfnamefont {Tatsufumi}\
  \bibnamefont {Nakamura}}, \bibinfo {author} {\bibfnamefont {Stepan~S.}\
  \bibnamefont {Bulanov}}, \bibinfo {author} {\bibfnamefont {Alexei~G.}\
  \bibnamefont {Zhidkov}}, \bibinfo {author} {\bibfnamefont {Yoshiaki}\
  \bibnamefont {Kato}}, \ and\ \bibinfo {author} {\bibfnamefont {Georg}\
  \bibnamefont {Korn}},\ }\bibfield  {title} {\enquote {\bibinfo {title} {{On
  extreme field limits in high power laser matter interactions: radiation
  dominant regimes in high intensity electromagnetic wave interaction with
  electrons}},}\ }in\ \href {\doibase 10.1117/12.2020847} {\emph {\bibinfo
  {booktitle} {High-Power, High-Energy, and High-Intensity Laser Technology;
  and Research Using Extreme Light: Entering New Frontiers with Petawatt-Class
  Lasers}}},\ Vol.\ \bibinfo {volume} {8780},\ \bibinfo {editor} {edited by\
  \bibinfo {editor} {\bibfnamefont {Georg}\ \bibnamefont {Korn}}, \bibinfo
  {editor} {\bibfnamefont {Luis~Oliveira}\ \bibnamefont {Silva}}, \ and\
  \bibinfo {editor} {\bibfnamefont {Joachim}\ \bibnamefont {Hein}}},\ \bibinfo
  {organization} {International Society for Optics and Photonics}\ (\bibinfo
  {publisher} {SPIE},\ \bibinfo {year} {2013})\ pp.\ \bibinfo {pages} {185 --
  199}\BibitemShut {NoStop}%
\bibitem [{\citenamefont {Neitz}\ and\ \citenamefont
  {Di~Piazza}(2013)}]{Neitz_2013}%
  \BibitemOpen
  \bibfield  {author} {\bibinfo {author} {\bibfnamefont {N.}~\bibnamefont
  {Neitz}}\ and\ \bibinfo {author} {\bibfnamefont {A.}~\bibnamefont
  {Di~Piazza}},\ }\bibfield  {title} {\enquote {\bibinfo {title} {Stochasticity
  effects in quantum radiation reaction},}\ }\href {\doibase
  10.1103/PhysRevLett.111.054802} {\bibfield  {journal} {\bibinfo  {journal}
  {Phys. Rev. Lett.}\ }\textbf {\bibinfo {volume} {111}},\ \bibinfo {pages}
  {054802} (\bibinfo {year} {2013})}\BibitemShut {NoStop}%
\bibitem [{\citenamefont {Vshivkov}\ \emph {et~al.}(1998)\citenamefont
  {Vshivkov}, \citenamefont {Naumova}, \citenamefont {M.}, \citenamefont
  {Pegoraro},\ and\ \citenamefont {Bulanov}}]{Vshivkov_1998}%
  \BibitemOpen
  \bibfield  {author} {\bibinfo {author} {\bibfnamefont {V.~A.}\ \bibnamefont
  {Vshivkov}}, \bibinfo {author} {\bibnamefont {Naumova}}, \bibinfo {author}
  {\bibfnamefont {N.}~\bibnamefont {M.}}, \bibinfo {author} {\bibfnamefont
  {F.}~\bibnamefont {Pegoraro}}, \ and\ \bibinfo {author} {\bibfnamefont
  {S.~V.}\ \bibnamefont {Bulanov}},\ }\bibfield  {title} {\enquote {\bibinfo
  {title} {Nonlinear electrodynamics of laser pulse interaction with a thin
  foil},}\ }\href@noop {} {\bibfield  {journal} {\bibinfo  {journal} {Phys.
  Plasmas}\ }\textbf {\bibinfo {volume} {5}},\ \bibinfo {pages} {2727}
  (\bibinfo {year} {1998})}\BibitemShut {NoStop}%
\bibitem [{\citenamefont {Bulanov}\ \emph {et~al.}(2016)\citenamefont
  {Bulanov}, \citenamefont {Esarey}, \citenamefont {Schroeder}, \citenamefont
  {Bulanov}, \citenamefont {Esirkepov}, \citenamefont {Kando}, \citenamefont
  {Pegoraro},\ and\ \citenamefont {Leemans}}]{Bulanov_2016}%
  \BibitemOpen
  \bibfield  {author} {\bibinfo {author} {\bibfnamefont {S.~S.}\ \bibnamefont
  {Bulanov}}, \bibinfo {author} {\bibfnamefont {E.}~\bibnamefont {Esarey}},
  \bibinfo {author} {\bibfnamefont {C.~B.}\ \bibnamefont {Schroeder}}, \bibinfo
  {author} {\bibfnamefont {S.~V.}\ \bibnamefont {Bulanov}}, \bibinfo {author}
  {\bibfnamefont {T.~Zh.}\ \bibnamefont {Esirkepov}}, \bibinfo {author}
  {\bibfnamefont {M.}~\bibnamefont {Kando}}, \bibinfo {author} {\bibfnamefont
  {F.}~\bibnamefont {Pegoraro}}, \ and\ \bibinfo {author} {\bibfnamefont
  {W.~P.}\ \bibnamefont {Leemans}},\ }\bibfield  {title} {\enquote {\bibinfo
  {title} {Radiation pressure acceleration: The factors limiting maximum
  attainable ion energy},}\ }\href@noop {} {\bibfield  {journal} {\bibinfo
  {journal} {Phys. Plasmas}\ }\textbf {\bibinfo {volume} {23}},\ \bibinfo
  {pages} {056703} (\bibinfo {year} {2016})}\BibitemShut {NoStop}%
\end{thebibliography}%

\end{document}